\documentclass[a4paper, amsfonts, amssymb, amsmath, reprint, showkeys, nofootinbib, twoside, natbib]{revtex4-1}
\usepackage[english]{babel}
\usepackage[utf8]{inputenc}
\usepackage[colorinlistoftodos, color=green!40, prependcaption]{todonotes}
\usepackage{lipsum}
\usepackage[dvipsnames]{xcolor}
\usepackage{xcolor}
\usepackage{booktabs}
\usepackage{adjustbox}
\usepackage{amsthm}
\usepackage{mathtools}
\usepackage{physics}
\usepackage{graphicx}
\usepackage[left=23mm,right=13mm,top=35mm,columnsep=15pt]{geometry} 
\usepackage{adjustbox}
\usepackage{placeins}
\usepackage{physics}
\usepackage[T1]{fontenc}
\usepackage{lipsum}
\usepackage{multirow}
\usepackage{csquotes}
\usepackage{caption}
\usepackage{subcaption}
\usepackage{float}
\usepackage{afterpage}

\usepackage[pdftex, pdftitle={A multichannel generalization of the HAVOK method}, pdfauthor={Carlos Colchero, Jorge Pérez, Alvaro Herrera and Oliver Probst}]{hyperref} 
\hypersetup{
    colorlinks=true,
    linkcolor=blue,  
    citecolor=blue,
    urlcolor=black
}

\begin{document}

\makeatletter
\def\@hangfrom@section#1#2{\@hangfrom{#1}{#2}} % Sentence case in section titles 
\makeatother

\title{A multichannel generalization of the HAVOK method for the analysis of nonlinear dynamical systems}

\author{Carlos Colchero} 
\email[]{A00835489@tec.mx}

\author{Jorge Perez}
\email[]{A00835537@tec.mx}

\author{Alvaro Herrera}
\email[]{A00836274@tec.mx}

\author{Oliver Probst}
\email[]{oprobst@tec.mx}

\affiliation{School of Engineering and Sciences, Tecnológico de Monterrey, Av. Eugenio Garza Sada 2051 Sur, Monterrey CP 64700, Mexico}

\begin{abstract}

By extending Takens' embedding theorem (1981), Deyle and Sugihara (2011) provided a theoretical justification for using parallel measurement time series to reconstruct a system's attractor. Building on Takens' framework, Brunton et al. (2017) introduced the Hankel alternative view of Koopman (HAVOK) algorithm, a data-driven approach capable of linearizing chaotic systems through delay embeddings. In this work, a modified version of the original algorithm is presented (\emph{m}HAVOK), a practical realization of Deyle and Sugihara's generalized embedding theory. \emph{m}HAVOK extends the original algorithm from one to multiple input time series and introduces a systematic approach to separating linear and nonlinear terms. An $R^2$-informed quality score is introduced and shown to be a reliable guide for the selection of the reduced rank. The algorithm is tested on the familiar Lorenz system, as well as the more sophisticated Sprott system, which features different behaviors depending on the initial conditions. The quality of the reconstructions is assessed with the Chamfer distance, validating how \emph{m}HAVOK allows for a more accurate reconstruction of the system dynamics. The new methodology generalizes HAVOK by allowing the analysis of multivariate time series, fundamental in real life data-driven applications.

% A modified version (\emph{m}HAVOK) of the Hankel alternative view of Koopman (HAVOK) algorithm introduced by Brunton et al. (2017) is presented. \emph{m}HAVOK generalizes the original approach from one to multiple input time series and introduces a systematic approach to separating linear and nonlinear terms. The results are demonstrated on the familiar Lorenz system, as well as the more complex Sprott system. \emph{m}HAVOK is shown to allow for a more accurate reconstruction of the system dynamics, particularly in the highly non-trivial case of the Sprott system. A two-stage fitting approach is used to first distinguish linear from nonlinear components for a fixed arbitrary reduced rank $r$ of the SVD-transformed block-Hankel matrix, followed by a refinement stage to extract the linear or core dynamics of the system. The quality of the attractor reconstruction was assessed with the Chamfer and the Hausdorff distances. An $R^2$-informed, $r$-dependent quality score is introduced and shown to be a good guide for the selection of the reduced rank, particularly in the case of the Sprott system, where the solution space is known to divide into two separate objects. The quality score is shown to be a good predictor of the accuracy of the reconstruction. The new methodology is believed to be useful for a wider use of the HAVOK approach, particularly its application to data-driven problems.

\end{abstract}

\date{\today} 
\maketitle
\makeatletter
\def\ps@titlepage{
  \def\@oddhead{} 
  \def\@evenhead{}
  \def\@oddfoot{\hfill\thepage\hfill} 
  \def\@evenfoot{\hfill\thepage\hfill}
}

\pagestyle{titlepage}
\makeatother

\section*{Glossary} \label{sec:glossary}

\begin{table}[H]
\centering
\begin{adjustbox}{scale=0.68} 
    \renewcommand{\arraystretch}{1.5}
    \setlength{\tabcolsep}{10pt}
    \begin{tabular}{rl}
            HAVOK & Hankel Alternative view of Koopman \\
            mHAVOK & multiple Hankel Alternative view of Koopman \\
            %DMD & Dynamical Mode Decomposition \\
            SVD & Singular Value Decomposition \\
            $\{\textbf{x}_n\}_{n=0}^{G-1}$ &  $G$ state variables \\
            $M$ & embedding dimension \\
            $f(\textbf{x}_n)$ & observable \\
            $\Psi[n]$ & discrete delay vector \\
            $\Psi_{GL}$ & multichannel delay vector \\ 
            $\tau_s$ & user-defined delay \\
            %RMSE & Root Mean Squared Error \\
            $\mathbf{h}_k$ & block of observables \\
            $\mathbf{H}$ & block Hankel matrix \\
            $N$ & size of the time series \\
            $Q$ & number of channels \\
            $l$ & $\min(QM, N-M-1)$ \\
            $\mathbf{U}_l$ & left singular vectors \\
            $\mathbf{S}_l$ & diagonal matrix of singular values \\
            $\mathcal{V}_l^T$ & eigen-time-delay coordinates \\
            $r$ & cutoff rank of $\mathcal{V}_l^T$ \\
            $\mathbf{V}$ & matrix with first $r$ columns of $\mathcal{V}_l^T$ \\
            $\tau$ & user-defined threshold \\
            $\dot{\textbf{V}}$ & derivative of the dynamical modes \\
            $\mathbf{\dot{v}}_j$ & \textit{j}-th column of $\mathbf{\dot{V}}$ \\
            $I_c$ & index set of core dimensions \\
            $I_f$ & index set of forcing dimensions \\
            $\mathbf{V}_c$ & $\mathbf{V}$ matrix with the core dimensions \\
            $\mathbf{V}_f$ & $\mathbf{V}$ matrix with the forcing dimensions \\
            $\mathbf{A}_r$ & Matrix describing the evolution of core dimensions \\

    \end{tabular}
\end{adjustbox}
\end{table}

\begin{table}[H]
\centering
\begin{adjustbox}{scale=0.68} 
    \renewcommand{\arraystretch}{1.5}
    \setlength{\tabcolsep}{10pt}
    \begin{tabular}{rl}
          
            $\mathbf{a}_j$ & \textit{j}-th column of $\textbf{A}$\\
            ${{\bar{\dot{v}}}_i}$ & temporal mean of $\mathbf{\dot{v}}_i$\\
            $\mathbf{\hat{A}}_r$ & system matrix \\
            $\mathbf{B}_r$ & input matrix \\
            $\mathbf{u}(t)$ & $\mathbf{V}_f$ \\
            $\mathbf{C}$ & output matrix \\
            $\mathbf{D}$ & zero matrix \\
            $\mathbf{x}(t)$ & simulated state vector \\
            $\mathbf{y}(t)$ & simulated output \\
            $\mathbf{x}(0)$ & first row of $\textbf{V}_{c}$ \\
            %$\textbf{y}(t)$ & Simulated matrix \\
            $N_s$ & simulated time steps \\
            $\mathbf{\hat{H}}(t)$ & simulated Hankel matrix \\
            $\textbf{U}_c$ & core left singular vectors \\
            $\mathbf{\Sigma}_c$ & diagonal matrix of core singular values\\
            $\cdot^{(\text{tr})}$ & Matrix employed on the training set \\
            $\cdot^{(\text{ev})}$ & Matrix employed on the evaluation set \\
            $\sigma_k$ & \textit{k}-th singular value \\
            $q$ & percentile cutoff \\
            $\mathcal{Q}(r)$ & quality score \\
            $C$ & subset of rank-values \\
            $\kappa(r)$ & condition number of $\mathbf{B}_r$\\      
            ${R}_q^2$ & $R^2$ between the \textit{q}-th channel of \\
            & the original and the reconstructed system\\
            ${\overline{R_{rec}^2}(r)}$ & mean of ${R}_q^2$ \\
            $\mathbb{V}$ & explained variance \\
            T & simulated time \\
            $\Delta t$ & time step \\
            $\Omega$ & Set of subsampled points \\
            $\zeta$ & Set of constructed points \\
            $d_c(\Omega,\zeta)$ & Chamfer distance between $\Omega$ and $\zeta$ \\
    
    \end{tabular}
\end{adjustbox}
\end{table}

 % $d_H(\Omega,\zeta)$ & Hausdorff distance between $\Omega$ and $\zeta$ \\
 %SINDy & Sparse identification of nonlinear dynamics \\
 %$\mathcal{F}(r)$ & Frobenius ratio \\

\newpage
\section{Introduction} \label{sec:introduction}

\raggedbottom

The linearization of chaotic dynamical systems has attracted considerable attention in recent years \cite{linearization_chaotic_dynamical_systems_reference1, linearization_chaotic_dynamical_systems_reference2, linearization_chaotic_dynamical_systems_reference3, linearization_chaotic_dynamical_systems_reference4, linearization_chaotic_dynamical_systems_reference5}. A great deal of this work can be traced back to a seminal paper by Bernard Koopman in 1931 \cite{Koopman_1931}, who proposed a Hilbert space description of classical mechanics, enabling the analysis of nonlinear systems through a linear operator. This operator, later known as the Koopman operator, acts on functions of the state space, also known as observables, by evolving them linearly in time, even when the system's evolution is nonlinear \cite{Mezic_Says_LinearizingThroughKoopman}. With the rise of modern data analysis, particularly for data-driven analysis of dynamical systems, Koopman's theory has seen a new surge in popularity \cite{data_driven_and_koopman_reference1, data_driven_and_koopman_reference2, data_driven_and_koopman_reference3, data_driven_and_koopman_reference4}.

A number of works \cite{work_inspired_by_mezic_reference1, work_inspired_by_mezic_reference2, work_inspired_by_mezic_reference3, rowley2009spectral, continuous_spectrum_reference1, continuous_spectrum_reference3, Korda_Kernels} have been sparked by pioneering work conducted by Igor Mezic and co-workers on the spectral analysis of the Koopman operator. Mezic demonstrated that, under ergodic conditions, the Koopman operator is unitary, allowing for the decomposition of its spectrum into a discrete and a continuous part \cite{Mezic2005}. The discrete part is associated with an almost-periodic and recurrent behavior, such as rotations on a torus, periodic orbits or coherent structures \cite{discrete_spectrum}. On the other hand, the continuous spectrum captures dispersive and mixing behavior exhibited by chaotic attractors or ergodic systems that lack quasi-periodic components \cite{koopman_spectrum_description}. In general, such dynamical systems exhibit a mixture of both discrete and continuous spectral components.

Building on this perspective, Brunton et al. \cite{Brunton_Havok_Nature} formulated the Hankel alternative view of Koopman (HAVOK) algorithm, a data-driven approach capable of linearizing chaotic systems. While not explicitly based on Koopman spectral theory, HAVOK reflects a similar approach by decomposing the dynamics into recurrent, structured are associated with the discrete spectrum, and irregular components that resemble continuous or mixed spectral behavior. In this algorithm, a delay embedding is constructed from a single time series and arranged into a Hankel matrix. Theoretically, this construction generates a Krylov subspace of observables \cite{Mezic_Proves_DMD_Relates_to_Koopman}, which approximates the action of the Koopman operator. Then, a Singular Value Decomposition (SVD) is performed, thereby identifying dynamical modes within the embedding. Only the leading dynamical modes are selected for further processing, requiring the definition of a cutoff rank $r$. The most suitable value of $r$ was determined empirically. The last remaining mode is interpreted as a forcing component driving the chaotic behavior of the system. Subsequently, a linear regression between the time derivatives of the first $r-1$ dynamical modes and the modes themselves is performed, leading to the construction of a dynamical matrix. Additionally, a second regression is carried out between the same derivatives and the last mode, allowing for the identification of a forcing matrix. Finally, both the dynamical and forcing matrices are combined into a linear system, from which time series can be simulated.

Although the algorithm was shown to successfully reconstruct several chaotic attractors, a number of limitations have become apparent \cite{havok_limitations_1, havok_limitations2, havok_limitations3}. First and foremost, HAVOK is limited to single-input single-output systems, precluding the use of multiple measurement time series. The algorithm also assumes that only a single nonlinear term provides the forcing of the system, which is taken as the last column of the reduced rank embedding, but no objective criterion for this separation approach was offered. Given the selection of the cutoff rank also remained arbitrary, it was unclear if HAVOK can be used for systems with unknown dynamics. Furthermore, another somewhat more subtle issue with HAVOK is the lack of a clear instruction for inverting the output time series from the eigen-time-delay-coordinate frame back to the original coordinates.

The use of a single time series for the reconstruction of the system dynamics in HAVOK is based on Takens' well-known embedding theorem \cite{Takens_Embedding_Theorem}. Building on earlier work by Packard et al. \cite{Packard_1980} and extended to fractal sets by Sauer et al. \cite{Sauer1991}, this theorem guarantees a diffeomorphic reconstruction of an attractor from a single observable given enough data points. Nevertheless, there
may be practical limitations to this approach. Deyle and Sugihara \cite{takens_theorem_generalized} proved generalized embedding theorems, showing that state-space reconstruction can be improved by including multiple measurement time series, thereby generalizing Takens' original result. Importantly, they proved that the reconstructed system remains diffeomorphic to the original attractor. This approach reduces the number of delays required from any time series, making the reconstruction less reliant on a single source of information. Such improvement is crucial, since Letellier and Aguirre \cite{Letellier2002} showed that if an observable remains invariant under a symmetry transformation, it will fail to unfold the system's topology when an embedding is performed on it, leading to a symmetry-blind reconstruction. Hence, single-input methods like HAVOK will fail to reconstruct the system if the observable is not properly selected. 
%or if it is the only one available from experimental data. 

In this work, the limitations of the HAVOK method pointed out above are addressed. Firstly, the method is extended to allow for multiple measurement time series to be processed together in block-Hankel matrices, providing a practical implementation of Deyle and Sugihara's work \cite{takens_theorem_generalized}. This extension will be referred to as \emph{m}HAVOK. The convenience of using multichannel information, albeit in the context of Dynamical Mode Decomposition (DMD), was already demonstrated by Arbabi and Mezic \cite{Mezic_Proves_DMD_Relates_to_Koopman}, and the use of block-Hankel matrices was proposed as early as 1985 by Juang and Pappa \cite{ERA_algorithm} for modal parameter estimation and model reduction. However, to the best knowledge of the authors, no multichannel version of the HAVOK algorithm has been presented so far. It will be shown that \emph{m}HAVOK allows for an improved reconstruction of the system dynamics, preventing symmetry-blindness reconstructions. Furthermore, a natural way of tracing the system trajectories in the original coordinate space is included, as well as an objective criterion for selecting the cutoff rank of the leading dynamical modes.

Regarding the second limitation of the original HAVOK method, i.e., the selection of nonlinear components, a systematic approach to solving this problem is also included in \emph{m}HAVOK. Instead of assuming the last dynamical mode from the reduced rank SVD must be classified as nonlinear, \emph{m}HAVOK performs a regression-based technique in order to allow for the identification of nonlinear terms. It will be shown that, in general, several nonlinear terms emerge and their inclusion is critical for an optimal reconstruction of the system dynamics. Finally, the quality of reconstruction is assessed quantitatively using similarity metrics as opposed to previous work \cite{Brunton_Havok_Nature}, where only qualitative comparisons were provided.

It should be noted that \emph{m}HAVOK, very much like HAVOK, has been designed as a technique capable of reconstructing the dynamics of (generally nonlinear) systems from observed data alone, i.e., as a data-driven technique. However, in order to validate the accuracy of the method, reference cases are required. In this work, time series of observables obtained from two test systems, the Lorenz and the Sprott system, determined from numerical solutions of the constitutive equations, were used as the input to \emph{m}HAVOK. The Lorenz system, a widely used reference case, is used for the illustration of some basic insights and for comparison with the original (1D) HAVOK method. The more challenging Sprott system, on the other hand, allows the new method proposed in this work to demonstrate its full potential, which is why this study case is explored in a somewhat greater detail.

The paper is organized as follows: First, Sections~\ref{subsec:generalized_embedding}-\ref{subsec:reconstructed_attractor} explain the extension of the HAVOK method to multiple inputs, present a novel component-classification scheme for the dynamical modes, and describe the procedure for reconstructing the system in the original coordinate system. Next, Sections~\ref{subsec:rank_selection}-\ref{subsec:test_systems_studied} provide an objective criterion for selecting the cutoff rank $r$, explain how the accuracy of the system's reconstruction is quantitatively assessed, and show the simulation setup for the two tested systems: Lorenz and Sprott. Then, Sections~\ref{sec:results_lorenz} and \ref{sec:results_sprott} present the simulation results, showing how the aforementioned modifications remarkably improve state-space reconstruction, particularly in the challenging Sprott system. Finally, Section~\ref{sec:conclusions} summarizes the work by presenting the main contributions, limitations found, and future research lines.

\section{Methods and simulation setup} \label{sec:methodology}

\subsection{Generalized Embedding}\label{subsec:generalized_embedding}

%The original HAVOK framework uses delay-coordinate reconstruction for 
Consider a discrete trajectory of unobserved state variables $\{\textbf{x}_n\}_{n=0}^{G-1}$ evolving on a compact state space $\mathcal{M}$. Given a smooth function of the state variables $f:\mathcal{M}\rightarrow \mathbb{R}$, the discrete delay vector of length $M$ and delay $\tau_s \in \mathbb{N}$ is defined by

\begin{equation}
    \Psi[n]=(f(\textbf{x}_{n}),f(\textbf{x}_{n+{\tau_s}}),\cdots,f(\textbf{x}_{n+{[M-1]\tau_s}})).
    \label{eq:delayvector}
\end{equation}

In the literature \cite{Korda_Kernels, Mezic2005, discrete_spectrum, koopman_spectrum_description}, $f(\textbf{x}_{n})$ is referred to as an observable. Under Takens' hypotheses, the delay map in Eq.~\eqref{eq:delayvector} is an embedding for generic observables, so that $\Psi(\mathcal{M})$ is diffeomorphic to the original system. In HAVOK \cite{Brunton_Havok_Nature}, columns of such delay vectors are stacked to form a Hankel matrix with uniform delays. 

An observable $f$ is defined as symmetry-blind with respect to a system symmetry $\mathcal{G}$ if $f(\mathbf{x}_n)=f(\mathcal{G}\cdot \mathbf{x}_n)$ $ \forall \ \mathbf{x}_n\in \mathcal{M}$. If an observable satisfies this condition, it will fail to unfold the system's topology when an embedding is performed on it \cite{Letellier2002}.

%These observables are invariant under system symmetries, which can prevent single-channel delay maps from unfolding the attractor \cite{Letellier2002}. The multiple observable extension mitigates this, as will be later demonstrated.

The previously discussed single-channel limitations can be avoided by extending the delay map to multiple observables, an extension provided by Deyle and Sugihara's generalized embedding theorems \cite{takens_theorem_generalized}. Consider $Q$ observables denoted by $\{f_q\}_{q=0}^{Q-1}$. In each of them, a multichannel delay vector of $M$ delays is collected as follows:

%Given that an improper selection of the observable may fail to unfold the system's topology when an embedding is performed on it \cite{Letellier2002}, 

\begin{equation}
    \Psi_{GL}[n]=\left(f_0(\textbf{x}_{n+k\tau_s}),\cdots,f_{Q-1}(\textbf{x}_{n+k\tau_s})\right)_{k=0}^{M-1},
    \label{deylemap}
\end{equation}

where  $\Psi_{GL}$ has the length $MQ$. For a fixed embedding dimension $D$, one may trade depth and channels so that $D=MQ$, an improvement from the single source delay, where $D=M$. With this considered, a {block Hankel matrix} is formed considering $\tau_s=1$, similar to the original formulation by Brunton et al. \cite{Brunton_Havok_Nature}. For each delay $k=\{0,\cdots,N-1\}$, the stacked observable block is given by:

\begin{align}
    \mathbf{h}_k := 
\begin{bmatrix}
    f_0(\textbf{x}_{k}) \\
    f_1(\textbf{x}_{k}) \\
    \vdots \\
    f_{Q-1}(\textbf{x}_{k})
\end{bmatrix}
\in \mathbb{R}^Q. \label{eq:observable_vector}
\end{align}

Arranging these blocks into the Hankel matrix yields:

\begin{align}
    \mathbf{H} = 
\begin{bmatrix}
    \mathbf{h}_0 & \mathbf{h}_1 & \cdots & \mathbf{h}_{N-M} \\
    \mathbf{h}_1 & \mathbf{h}_2 & \cdots & \mathbf{h}_{N-M+1} \\
    \vdots      & \vdots      & \ddots & \vdots            \\
    \mathbf{h}_{M-1} & \mathbf{h}_{M} & \cdots & \mathbf{h}_{N-1}
\end{bmatrix}
,\label{eq:block_hankel_matrix_generalized} % \in \mathbb{R}^{QM\times (N-M+1)}
\end{align}

where $N$ is the number of time measurements, i.e., the size of the time series. Each column within the block Hankel matrix constitutes a time-delay vector $\Psi[n]$, such as the one presented in Eq.~\eqref{eq:delayvector}.

The array in Eq.~\eqref{eq:block_hankel_matrix_generalized} extends the delay embedding to several input dimensions,
with the matrix size being $(QM, N-M+1)$. This embedding allows for several spatiotemporally correlated channels to coexist, potentially enriching the system's reconstruction by reducing noise correlations occurring in one-dimensional embeddings \cite{takens_theorem_generalized}. 

Having constructed the block Hankel matrix, a thin Singular Value Decomposition (SVD) is performed \cite{thin_SVD},

\begin{align}
\label{SVD}
    \mathbf{H} = \mathbf{U}_l\mathbf{S}_l\mathbf{\mathcal{V}}_l^T,
\end{align}

where $l =\min(QM,N-M+1)$. Similar to the original HAVOK method \cite{Brunton_Havok_Nature}, this decomposition identifies an orthonormal basis for the dynamical modes of the system. In practical implementations, singular values are usually ordered from largest to smallest, as well as their associated vectors in $\mathcal{V}_l$ and $\textbf{U}_l$. This ordering affects the indexing of the SVD modes, but it does not permute the rows or columns of the original matrix \textbf{H}. Therefore, channel ordering and block positions in the original matrix are preserved after an SVD, regardless of the sorting procedure.

$\mathbf{U}_l$ is an orthonormal matrix sized $(QM, QM)$, whose columns form a basis for the column space of \textbf{H}. These columns are often referred to as spatial or structural modes, since they capture coherent patterns across data rows. In the context of Hankel matrices, each row is a shifted time window, so the spatial modes capture coherent structures within different time windows. This motivates the use of multiple channels in the embedding process, revealing coherent patterns across different measurement channels. Accordingly, using a block array enhances the reconstruction quality and better captures the global dynamics of the system.

On the other hand, $\mathbf{\mathcal{{V}}}_l^T$ is a rectangular matrix sized $(QM, N-M+1)$, whose rows contain the time coefficients of the structural modes in $\mathbf{U}_l$. For that reason, the rows of $\mathbf{S}_l\mathbf{\mathcal{V}}_l^T$ are often referred to as the eigen-time-delayed coordinates, since they serve as a transformed time series corresponding to the primary structural modes in the Hankel matrix.

\subsection{Linear regression in embedded space}\label{sub_sec:reg}

%\emph{m}HAVOK aims to reconstruct $\mathcal{V}_l^T$ through a learned linear model of the nonlinear system's temporal coefficients. 
Having constructed the generalized embedding, the eigen-time-delay coordinates are truncated by retaining the first $r$ columns of $\mathbf{\mathcal{V}}_l$. Let $\textbf{V}$ represent the reduced rank matrix. $r$ plays a fundamental role in the original HAVOK model, since the $r$-th column of $\textbf{V}$ is assumed to be the forcing term of the linearized model \cite{Brunton_Havok_Nature}. Given that the appropriate value of $r$ is not known a priori, a systematic approach is required. 

% Furthermore, it is not evident why the forcing term should be limited to the last dimension.

% To address these questions, a regression-classification approach is proposed. In this procedure, the $V_r$ components are classified based on a goodness-of-fit criterion, allowing for the identification of several nonlinear components.

To address these questions, each of the columns in $\mathbf{V}$ are classified based on a goodness-of-fit criterion, allowing for the identification of nonlinear components. As will be shown below, the number of nonlinear terms is generally not limited to one, and the inclusion of all nonlinear components meeting an appropriate selection criterion is critical to the correct reconstruction of the system dynamics.

A first regression problem is solved to separate linear and nonlinear components at a fixed rank $r$. The first step involves the calculation of the derivative of the eigen-time-delay coordinates $\mathbf{\dot{V}}$ using a high-order central difference, which enhances numerical accuracy and reduces sensitivity to noise. Then, a linear regression is performed in an attempt to find $\mathbf{A}_r$, i.e., the matrix that best describes the changes in time of the eigen-time-delay coordinates:

\begin{align}
    \min_{\mathbf{A}_r} ||\mathbf{\dot{V}}-\textbf{V}\mathbf{A_r}^T||_F, \label{eq:minimization} 
\end{align}

where $||\cdot||_F$ denotes the Frobenius norm. Therefore, each column of the time derivative matrix $\mathbf{\dot{V}}$ is approximated as

\begin{align}
    \mathbf{\dot{v}}_j\approx \textbf{V}\mathbf{a}_j,
\end{align}

where $\mathbf{\dot{v}}_j$ and $\mathbf{a_j}$ are the \textit{j}-th columns of $\mathbf{\dot{V}}$ and $\mathbf{A}_r$, respectively. Next, the coefficient of determination $R^2$ is calculated for each of the regressed columns,

\begin{align}
\label{r_squared}
    R_i^2=1-\frac{||\mathbf{\dot{v}}_i-\textbf{V}\textbf{a}_i||^2}{||\mathbf{\dot{v}}_i-\bar{\dot{{v}}}_i\cdot\textbf{1}||^2},
\end{align}

where $||\cdot||$ denotes the Euclidean norm in ${L}^2$, ${{\bar{\dot{v}}}_i}$ is the temporal mean of $\mathbf{\dot{v}}_i$ and $\textbf{1}$ is a vector of ones. Linear and nonlinear components are then classified by their goodness of fit:

\begin{align}
   \mathcal{I}_c&=\{i:R_i^2\ge\tau\}, \label{eq:user_threshold}\\
   \mathcal{I}_f&=\{1,\dots,r\} \backslash \mathcal{I}_c,
\end{align}

where $\tau$ is a user-defined threshold, $\mathcal{I}_c$ is the index set of core dimensions obtained with the given threshold and $\mathcal{I}_f$ is the index set for the identified nonlinear dimensions. Evidently, 

\begin{align}  
    r_c+r_f:=|\mathcal{I}_c|+|\mathcal{I}_f|=r,   
\end{align}

where $|\cdot|$ denotes the cardinality of the indexed sets. Since the two sets are disjoint, the cardinality of their union equals $r$. At this point, $r$ continues to be an arbitrary-fixed value. 

Next, the temporal coordinate matrix $\textbf{V}$ is redefined to only contain the core dimensions identified in the first regression stage:

\begin{align}
%\label{8}
    \textbf{V}_c := [\mathbf{v}_i]_{i\in\mathcal{I}_c},
    \label{VC}
\end{align}

where $\mathbf{v}_i$ is the \textit{i}-th column of $\mathbf{V}$. Therefore, $\textbf{V}_c$ only retains the columns that display a linear evolution. Similarly, it is possible to define a time series matrix that corresponds to nonlinear components:

\begin{align}
\label{9}
    \textbf{V}_f:=[\textbf{v}_i]_{i\in\mathcal{I}_f}.
\end{align}

To describe the time evolution of core dimensions (which show linear behavior), a second regression problem is defined. The objective is to obtain two matrices, one for the evolution of nonlinear terms, $\mathbf{B}_r$, and the other one for the evolution of core dimensions, $\mathbf{\hat{A}}_r$:

\begin{align}
\label{eq:second_reg}
    \min_{\mathbf{\hat{A}}_r,\mathbf{B}_r} ||\mathbf{\dot{V}}_c-\mathbf{V}_c \mathbf{\hat{A}}_r^T - \textbf{V}_f \textbf{B}_r^T||_F.
\end{align}

Subsequently, to simulate embedding dynamics, i.e., the dynamics of the eigen-time-delay coordinates, the retrieved system matrices are provided as input to the \textbf{lsim} function from Python's SciPy library, which simulates the following linear system in the familiar state-space form:

\begin{align}
\label{eq:linear_model}
\mathbf{\dot{x}}&=\hat{\mathbf{A}}_r\mathbf{x} + \mathbf{B}_r\mathbf{u}, \\
    \textbf{y} &= \textbf{C}\textbf{x}+\textbf{D}\mathbf{u},
\end{align}

where $\textbf{u}:=\textbf{V}_f$, $\textbf{C}$ is an $r_c \times r_c$ identity matrix, $\textbf{D}$ is a zero matrix, $\textbf{x}$ is the simulated state vector and $\textbf{y}$ is the simulated output. The initial condition $\textbf{x}(0)$ is a column vector given by the first row of $\mathbf{V}_c$, i.e.,

\begin{align}
    \textbf{x}(0) =\mathbf{V}_c(0,:).
\end{align}

With this procedure, the core eigen-time-delay coordinates are simulated by integrating the linear dynamical system defined by the matrices $\mathbf{\hat{A}}_r$ and $\textbf{B}_r$. The simulated matrix $\textbf{y}$ will have dimensions $(N_s,r_c)$, where $N_s\leq N$ denotes the number of simulated time steps. 

\subsection{Reconstructed Attractor}\label{subsec:reconstructed_attractor}

Having reconstructed the time series of the eigen-time-delay coordinates $\textbf{y}$, the simulated block Hankel matrix $\mathbf{\hat{H}}$ is retrieved by reapplying the SVD factorization:

\begin{align}
\label{inversetrans}
    \mathbf{\hat{H}}=\textbf{U}_c\mathbf{\Sigma}_c \textbf{y}^T \in \mathbb{R}^{QM\times (N-M+1)},
\end{align}

where $\textbf{U}_c$ and $\mathbf{\Sigma}_c$ represent the SVD matrices of the core components. As previously noted, the SVD factorization preserves channel ordering and block positions in the reconstructed block-Hankel matrix.

Given $Q\ge3$ observables, a three-dimensional representation of the embedded attractor can be obtained by plotting the columns from the first block of the simulated Hankel matrix. If $Q<3$, an embedded attractor in $M$-dimensional space is recovered. If the observable from which the embedding was obtained is smooth and generic, then it is diffeomorphic to the original attractor, according to Takens' theorem \cite{Takens_Embedding_Theorem}. Such attractor can be visualized in three-dimensional space (assuming the original attractor is three-dimensional). To do so, the columns from  three different row indices were chosen in a way that there is little correlation between them. 

In cases where a single observable was provided, the selected row indices were $k=0$, $k=M/2$ and $k=M$. This selection allowed for the visualization of the embedded attractor in three-dimensional space. If $Q=2$, only a single delay is required; in this case, a delay from any of the two inputted time series was chosen so that $k=M$.

In order to determine the goodness of the reconstruction of the system dynamics, the coefficient of determination between the original data set (assumed to have $Q$ input channels) and the reconstructed attractor (which features $Q$ reconstructed channels), has been calculated. Let $X^\text{rec}_{q,n}:=\mathbf{\hat{H}}(q,n)$, where $q\in\{0,\cdots,Q-1\}$, $q$ is the channel index (rows of a Hankel block) and $n$ is the time index for the reconstructed data series (columns of the Hankel matrix). When compared to the data series from the original Hankel matrix $X^\text{og}_{q,n}=\mathbf{H}(q,n)$, the goodness of fit is given by:

\begin{align}
    R_q^2=1-\frac{\sum_{\forall n}(X^\text{og}_{q,n}-X^\text{rec}_{q,n})^2}{\sum_{\forall n}(X_{q,n}^{\text{og}}-\bar{X}_{q}^{\text{rec}})^2},
\end{align}

where $\bar{X}_{q}^{rec}$ is the mean of $X_{q,n}^{\text{rec}}$ over the columns (time indices). Finally, a simple average is performed:

\begin{align}
    \overline{{R}_{rec}^2}:=\frac{1}{Q}\sum_{q=0}^{Q-1}R^2_q.
    \label{ReconstructedVsOriginal}
\end{align}

\subsection{Automated rank selection algorithm}\label{subsec:rank_selection}

In Section~\ref{sub_sec:reg} we explained how linear and nonlinear components in eigen-delay space can be distinguished by a certain regression procedure, once a value of the cutoff rank $r$ had been selected. However, nothing was said about how to identify such rank $r$. In Ref.~\cite{Brunton_Havok_Nature}, the authors found that, at least in the case of the Lorenz System, the precise value of $r$ was not critical for a good reconstruction of the system dynamics. For \emph{m}HAVOK, particularly in systems other than the Lorenz system, reconstruction quality changes drastically for different rank choices. Therefore, an appropriate scheme for the optimal selection of $r$ had to be devised. In the following, two schemes will be described: a variance-informed $r$ selection, used as a reference case, and an $\overline{R_\text{rec}^2}$-informed quality score, one of the innovations of this work.

\subsubsection{\texorpdfstring{Variance-informed $r$ selection}{Variance-informed r selection}}
\label{sub_sec:variance_rank_selection_attempt}

%In the supplementary information of \cite{Brunton_Havok_Nature}, Brunton et al. suggested the rank selection possess an intrinsic relation with the choice of singular value thresholding of the Hankel matrix. 

Given that both HAVOK and \emph{m}HAVOK are based on SVD techniques, it seems plausible to use a criterion involving the explained variance for the selection of the cutoff rank \cite{Analysis_SingularValues_HankelMatrix,
Optimal_Hard_Threshold_SingularValues}. This approach has been implemented by several authors \cite{chinos_HAVOK, HAVOK_In_Psychology_m_value}. It should be noted that Ref.~\cite{Optimal_Hard_Threshold_SingularValues} has a more sophisticated approach to cutoff rank selection, focusing on signal-to-noise ratio, something that has been stressed by Brunton et al. \cite{Brunton_Havok_Nature}.

%Even in 2024, this approach has continued to be implemented by multiple authors . This technique is frequently applied for SVD truncationbased on explained variance, which makes it a natural candidate for rank selection in HAVOK 

The reference method implemented in this work is based on the following: given the hierarchically sorted SVD singular values from Eq.~\eqref{SVD}, the explained variance for a given number of modes is calculated. Taking this into account, a cutoff rank is defined such that a given threshold of explained variance $\mathbb{V}$ is achieved,

\begin{align}
    \mathbb{V}=\frac{\sum_{k=1}^r\sigma_k^2}{\sum_{k=1}^{MQ}\sigma_k^2}\ (100\%), \label{eq:variance_rank_selection_procedure}
\end{align}

where $\sigma_k$ is the \textit{k}-th singular value. As previously mentioned, this method is only employed as a reference and does not constitute an original contribution of this work.

\subsubsection{\texorpdfstring{$\mathrm{R}^2$-informed quality score}{R²-informed quality score}} \label{sub_sec:quality_score}

The novel cutoff rank selection method proposed in this work is based on the dynamical matrix $\mathbf{B}_r$. This matrix is retrieved for a range of selected ranks $r\in[r_{\text{min}},r_\text{a}]$, where $r_{\text{min}}$ is the smallest rank for which \emph{m}HAVOK successfully identifies at least one linear mode and $r_\text{a}$ is an arbitrary upper bound. For each $\mathbf{B}_r$, a singular value decomposition is performed, 

\begin{equation}
    \mathbf{B}_r=U\Sigma V^{T}.
\end{equation}

Then, the quotient between the maximum and minimum singular values is stored for each $r\in[r_{\text{min}},r_\text{a}]$. Mathematically, 

\begin{equation}
 \kappa(r):=\sigma_\text{max}(\mathbf{B}_r)/\sigma_\text{min}(\mathbf{B}_r).
\end{equation}

The rationale behind this criterion stems from the condition number of a matrix \cite{condition_number_matrix}, which is defined as the quotient between the maximum and minimum singular values and measures how sensible a matrix is to perturbations in its input. In practice, $r$ values associated with a high quotient enable the system's linearization in embedded space. 

With this considered, a subset of high scoring rank values is selected by introducing a percentile cutoff. The upper $q\%$ of retrieved scores are stored in a subset $C\subseteq[r_{\text{min}},r_\text{a}]$, while the remaining $(100-q)\%$ are removed. For each $r\in C$, a $70/30$ training-testing split is performed on the (possibly multiple) measurement time series. Working on the training set, the routine determines the matrices $\mathbf{\hat{A}}_r$ and $\textbf{B}_r$ from the linear system, as well as the index sets $\mathcal{I}_c$ and $\mathcal{I}_f$. The nonlinear terms are obtained from a projection of the training set onto the evaluation set \cite{Brunton_Havok_Nature}:

\begin{align}\mathcal{V}_l^{(\text{ev})}=\mathbf{S}^{-1(\text{tr})}_l\mathbf{U}_l^{T(\text{tr})} \mathbf{H}^{(\text{ev})},
\end{align}

where $\cdot^{(\text{tr})}$ and $\cdot^{(\text{ev})}$ are matrices related to the training and evaluation sets, respectively. Having obtained the eigen-time-delay coordinates for the evaluation data, the nonlinear terms are given by:

\begin{align}
    \textbf{V}_f^{(\text{ev})}=[\textbf{v}^{(\text{ev})}_i]_{i\in \mathcal{I}_f^{(\text{tr})}}
\end{align}

where $\textbf{v}^{(\text{ev})}_i$ are columns of $\mathcal{V}_k^{(\text{ev})}$. Using the projected nonlinear terms and the dynamical matrices of the training period, the reconstructed time series for the evaluation set are simulated. Then, the mean $\overline{R_\text{rec}^2}$ defined in Eq.~\eqref{ReconstructedVsOriginal} is calculated between the original and the reconstructed evaluation time series after reapplying the SVD factorization. 

Each $r$-candidate in $C$ is scored on its ability to reconstruct the evaluation data set, which is provided by the following quality score:

\begin{align}
\mathcal{Q}(r)&:=1-\overline{R_\text{rec}^2}(r), \label{eq:quality_score_bn} 
\end{align}

The optimal rank value $r_{\text{opt}}$ corresponds to the one with a minimal quality score, 

\begin{align}
    r_{\text{opt}}&=\arg \min_{r\in C}\mathcal{Q}(r).\label{eq:rank_selection_bn}
\end{align}

Equivalently, $\mathcal{Q}(r_{\text{opt}}) = \min[\mathcal{Q}(r)] \ \ \forall r\in C$. We posit that this methodology provides an optimal criterion for rank selection. In order to test the validity of the rank selection algorithm, an independent metric for the fidelity of the system's reconstruction is required.

\begin{figure*}[htbp]
    \centering
    \includegraphics[width=\textwidth]{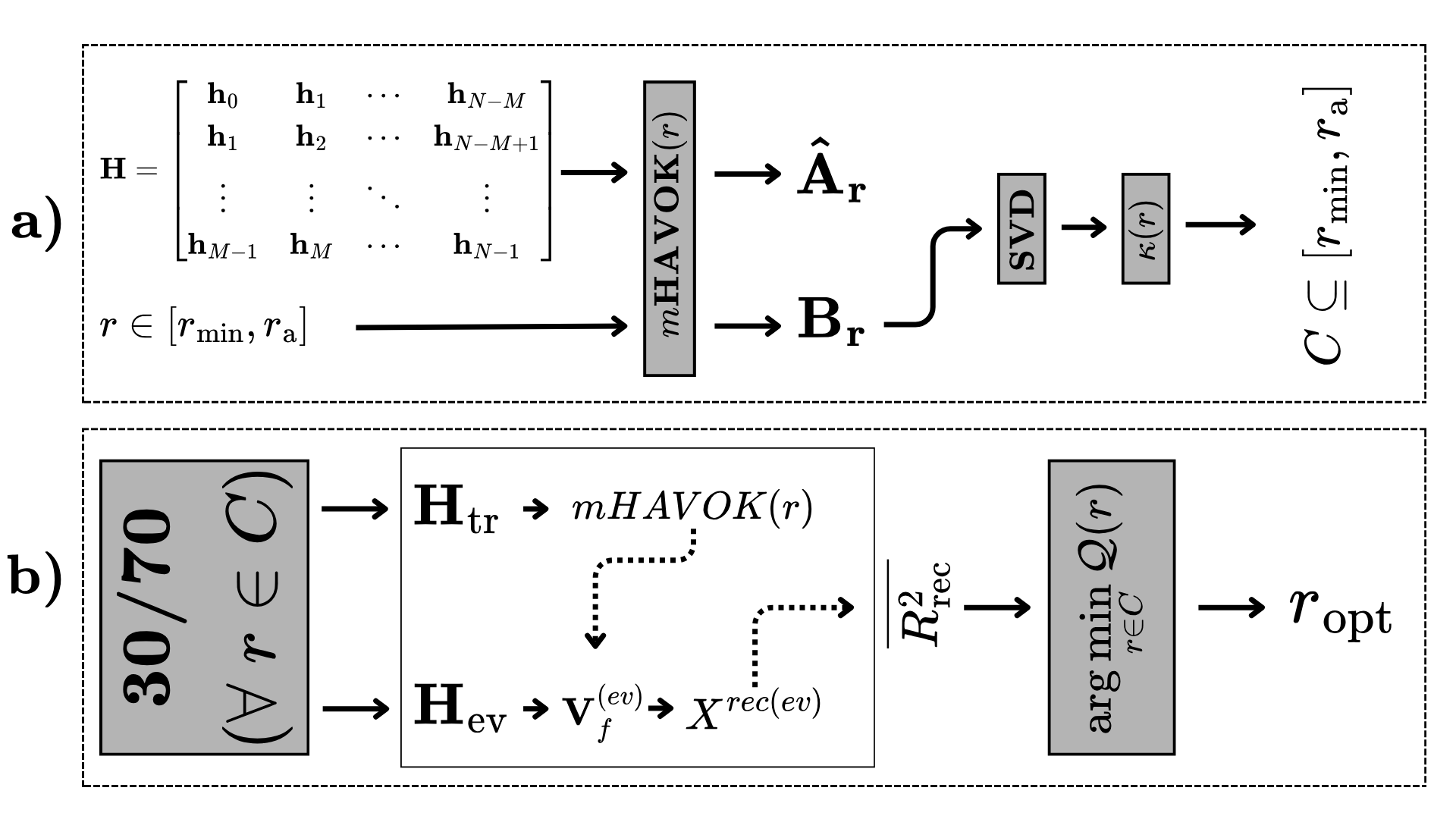} 
    \caption{General pipeline for the rank selection procedure developed and utilized throughout this work. The workflow starts in a), where a block Hankel matrix is built and a variety of cutoff ranks $r\in[r_\text{min}, r_\text{a}]$ are applied in order to obtain the subset $C$ of rank-candidates. In b), a condensed illustration of the 70/30 split procedure is presented, where the training set serves as input for \emph{m}HAVOK, and the evaluation set provides $\overline{\mathrm{R}_\text{rec}^2}$. Finally, the optimal rank $r_{\text{opt}}$ is defined as the one that minimizes the quality score $\mathcal{Q}(r)$.}
    \label{fig:diagrama_methodology}
\end{figure*}

A summary of this rank selection process, as well as the 70/30 training-testing split on the measurement time series, is presented on Fig.~\ref{fig:diagrama_methodology}.

\subsection{Quantitative assessment of the goodness of reconstruction} \label{sub_sec:geometric_criteria} 

In this work, the Chamfer distance \cite{chamfer_distance} was implemented to measure the accuracy of the system's reconstruction. This metric is calculated in the original coordinate space and provides a measure of the generalized distance between sets of points. Considering prior work on HAVOK only provided a descriptive reconstruction in eigen-delay space \cite{Brunton_Havok_Nature, chinos_HAVOK}, the current methodology builds on and goes beyond these works.

In our case, having applied the previously mentioned 70/30 split, we are interested in the distance between the simulated and reconstructed attractors for the evaluation data sets. To ensure computational efficiency, a random subsampling is first performed on the original index set $I_R=\{0,1,\ldots,N-1\}$, defining a new index set $I_S\subset I_R$ of size $S<N$. Given the generated matrices $X^\text{og}_{n,q}$ and $X^\text{rec}_{n,q}$ defined in Section~\ref{subsec:reconstructed_attractor}, two corresponding subsampled sets of points are defined: one for the points from the original attractor, 

\begin{equation}
    \Omega=\{ X^\text{og}_{n,:}:n\in I_S\},
\end{equation}

and one for the reconstructed attractor,

\begin{equation}
 \zeta=\{X^\text{rec}_{n,:}:n\in I_S\}. \label{eq:set_reconstructed}  
\end{equation}

Then, both data sets are rescaled by the min-max procedure \cite{min_max_normalization} to ensure repeatability and standardization, i.e., by setting $x_i\rightarrow (x_i - x_\mathrm{min})/(x_\mathrm{max} - x_\mathrm{min}) \ \forall x_i \in\zeta \land \Omega$. $x_\mathrm{max}$ and $ x_\mathrm{min}$ denote, respectively, the maximum and minimum values in each of the sets. These normalized subsampled sets serve as the basis for geometric validation. 

With this considered, the Chamfer distance $d_c$ quantifies the average closest-point distance between both sets; the smaller the distance, the higher the geometrical resemblance between the two sets of points. Mathematically, the Chamfer distance is defined as a function $d_C:  \mathbb{R}^{Q\times S} \times \mathbb{R}^{Q\times S} \rightarrow \mathbb{R}$, where $Q$ is the number of channels and $S$ is the subsampled size, such that

\begin{align}
    d_c(\Omega,\zeta) = \frac{1}{|\Omega|} \sum_{\omega \in \Omega} \min_{z \in \zeta} \|\omega - z\| \nonumber \\
    + \frac{1}{|\zeta|} \sum_{z \in \zeta} &\min_{\omega \in \Omega} \|z - \omega\|,
\end{align}

where $|\cdot|$ stands for the cardinality of the sets and $||\cdot||$ is the $Q$-dimensional Euclidean distance. In other words, for each $ \omega \in \Omega$, the minimum distance with respect to all points in $\zeta$ is calculated. All minimum distances are summed and then normalized. Similarly, a corresponding procedure is carried out for all points $z \in \zeta$. Finally, the Chamfer distance $d_c$ is calculated as the sum of the two partial distances. 

%Additionally, the proposed measure can be applied to state-space reconstructions obtained through delay embeddings. In such cases, the reconstructed attractor may appear rotated with respect to the original. If the Chamfer distance is computed at this stage, non-meaningful results would be obtained. To address this, one can multiply the reconstructed attractor by an orthogonal matrix that minimizes the distance to the original. This is a well-known optimization problem known as the orthogonal Procrustes problem, whose solution is given in \cite{Schönemann1966}. After this alignment, we claim the proposed distance can provide an objective measure of reconstruction quality even in multichannel comparison, although this extension is not pursued in the present work.

The Hausdorff distance \cite{Hausdorff_distance} was also employed to quantify the goodness of the reconstructions, but it showed similar results to those from the Chamfer distance; the corresponding results have therefore been omitted for brevity.

\subsection{Test systems studied}\label{subsec:test_systems_studied}

\subsubsection{Lorenz system}

The \emph{m}HAVOK method introduced in this work was first tested on the Lorenz system, which is given by

\begin{align} 
    \dot{x}&=\sigma(y-x),\label{eq:lorenz_system_x}\\
     \dot{y}&=x(\rho-z)-y,\label{eq:lorenz_system_y}\\
      \dot{z}&=xy-\beta z,\label{eq:lorenz_system_z}
\end{align}

where $\mathbf{r}=(x,y,z)^T \in R^3$ and the parameter values used throughout this work are

\[
\sigma = 10, \quad \rho = 28, \quad \beta = \frac{8}{3}.
\]

\begin{figure}[H]
    \centering
    \includegraphics[width=0.3\textwidth]{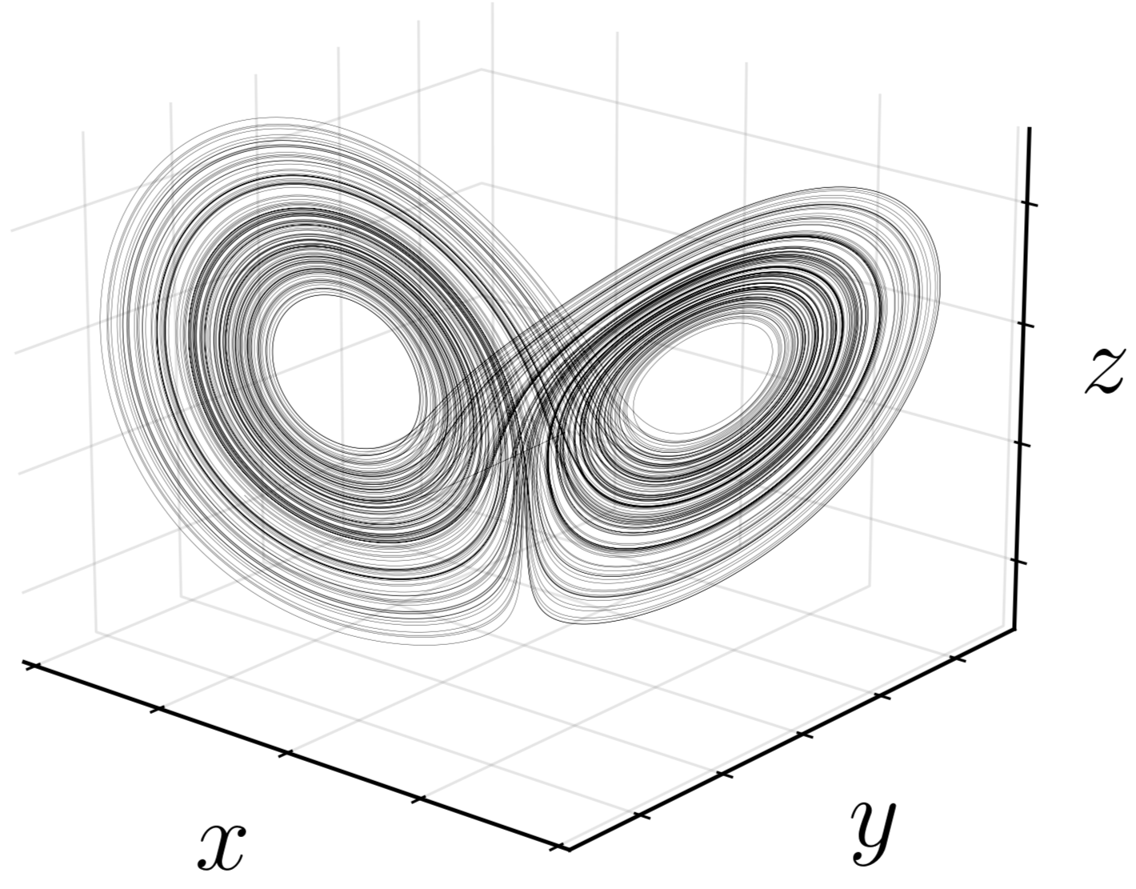} % elev=20, azim=130
    \caption{ Lorenz attractor generated using Eqs. \eqref{eq:lorenz_system_x}, \eqref{eq:lorenz_system_y} and \eqref{eq:lorenz_system_z}.}
    \label{fig:lorenz_original}
\end{figure}

The time step was chosen to be $\Delta t = 0.001$, while the selected simulation time was $T = 200$. With this configuration, the original Lorenz attractor shown in Fig.~\ref{fig:lorenz_original} was obtained. 

$Q$ channels were introduced into the embedding in Eq.\ \eqref{eq:block_hankel_matrix_generalized} as observables from the state variables $\textbf{x}=(x,y,z)$. The optimal rank value $r_\text{opt}$ was calculated using the methodology in Section\ \ref{sub_sec:quality_score}. The embedding dimension was set to $M = \lceil 100/Q \rceil$ per channel, where $\lceil\cdot\rceil$ denotes the ceiling function. Although this choice is empirical, it aligns with prior work in HAVOK and delay-coordinate embeddings for chaotic systems \cite{Brunton_Havok_Nature, HAVOK_In_Psychology_m_value}, where values of $M \in [100, 150]$ were shown to be sufficient to unfold the attractor's geometry. The Lorenz system was chosen because of its importance as a test case in dynamical systems modeling, as well as its central role in the original HAVOK paper \cite{Brunton_Havok_Nature}. Its role in this work is mainly to illustrate some of the qualitative improvements of \emph{m}HAVOK over HAVOK, such as its insensitivity to symmetry blindness and its capability of accurately reconstructing trajectories in original space. It is also used to illustrate the appearance of multiple nonlinear components and the importance of including them in the reconstruction procedure.

\subsubsection{Sprott system}

The algorithm's ability to reconstruct complex dynamical systems is tested through the exploration of the Sprott system, which is not globally ergodic \cite{SPROTT20141361}. This system is described by the following coupled differential equations:

\begin{align}
    \dot{x}&=y+2xy+xz, \label{eq:sprott_x} \\
    \dot{y}&=1-2x^2+yz, \label{eq:sprott_y} \\
    \dot{z}&=x-x^2-y^2. \label{eq:sprott_z} 
\end{align}

\begin{figure}[H]
    \centering
    \includegraphics[width=0.3\textwidth]{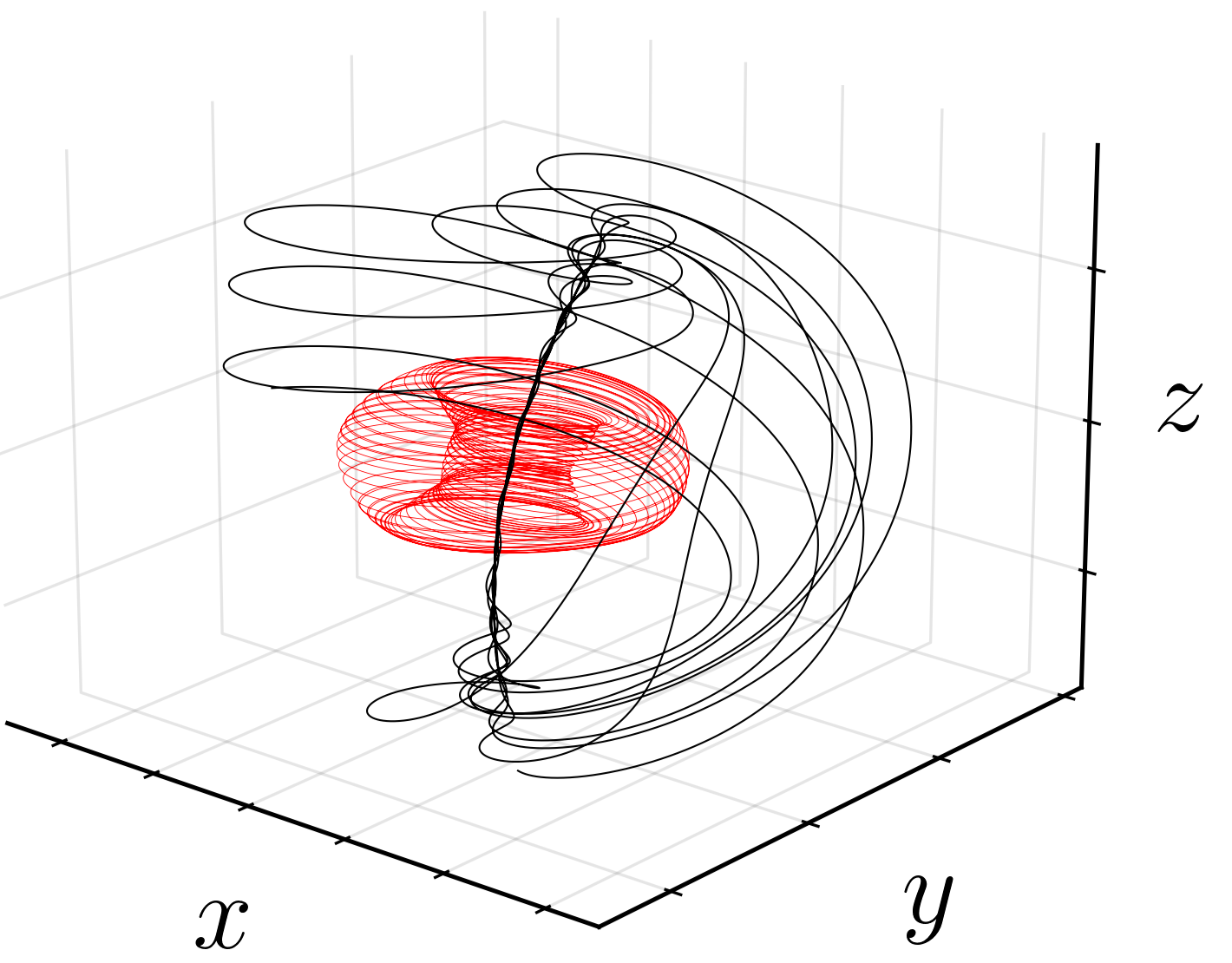}
    \caption{Sprott system generated using Eqs.\ \eqref{eq:sprott_x}, \eqref{eq:sprott_y} and \eqref{eq:sprott_z}. The initial conditions were $\textbf{r}_0=(1,0,0)^T$ (red) and $\textbf{r}_0'=(2,0,0)^T$ (black).}
    \label{fig:sprotts_original_attractor}
\end{figure}

Depending on the initial conditions $\mathbf{r}_0=(x_0,y_0,z_0)^T$, the system either converges to a strange attractor with a noninteger, fractal dimension, or resides in a non-attracting toroidal orbit, which is invariant under the system's flow \cite{ALGABA20191441}. Numerically, the embedding dimension was increased to $M = \lceil 500/Q \rceil$. The rest of the simulation parameters remained the same as in the Lorenz system; $T=200$ and $\Delta t=0.001$. Fig.~\ref{fig:sprotts_original_attractor} shows Sprott system for both types of initial conditions.

%The optimal $r$ is calculated using the previously presented methodology, while ensuring the total $QM\simeq100$ ($M=\text{ceil}(100/Q)$). 

% Additionally, the algorithm was executed multiple times for different values of $r\in[5,25]$ while keeping the embedding dimension fixed at $M = 35$ per channel. For each run, the previously defined Quality Score $\mathcal{Q}(r)$ was calculated, identifying the optimal $r$-regions for the Lorenz system, as presented in section \ref{sub_sec:rank}. In this case, a rank value $r=17$ was identified and selected.

% The current framework should be capable of identifying the appropriate rank for each set of initial conditions by calculating the Quality Score in Eq. \eqref{eq:quality_score}. Furthermore, by providing sufficient channel information, it should be able to accurately reconstruct both states of the system. Finally, it should successfully classify different nonlinear dimensions, separating linear components, potential forcing components and spurious modes with no previous assumptions.
 
% Measurements from the three observables $(x, y, z)$ were performed uniformly and collected into a block Hankel matrix, as described in Eq. \eqref{eq:Block_Hankel_Matrix}. 

\section{Results - Lorenz system} \label{sec:results_lorenz}

\begin{figure*}[htbp]
  \centering
  \begin{subfigure}{0.3\textwidth}
    \includegraphics[width=\linewidth]{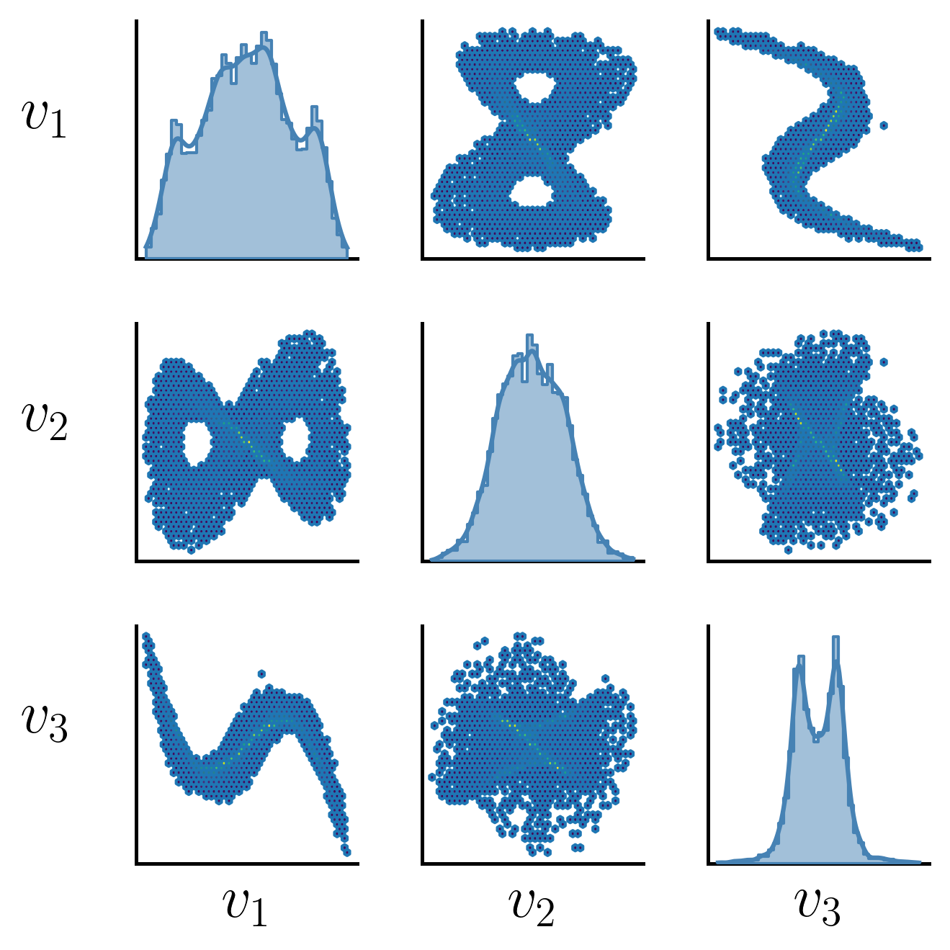}
    \caption{One-dimensional embedding provided by the $x_n$ channel.}
    \label{fig:sub1_bn}
  \end{subfigure}
  \hspace{2em}
  \begin{subfigure}{0.3\textwidth}
    \includegraphics[width=\linewidth]{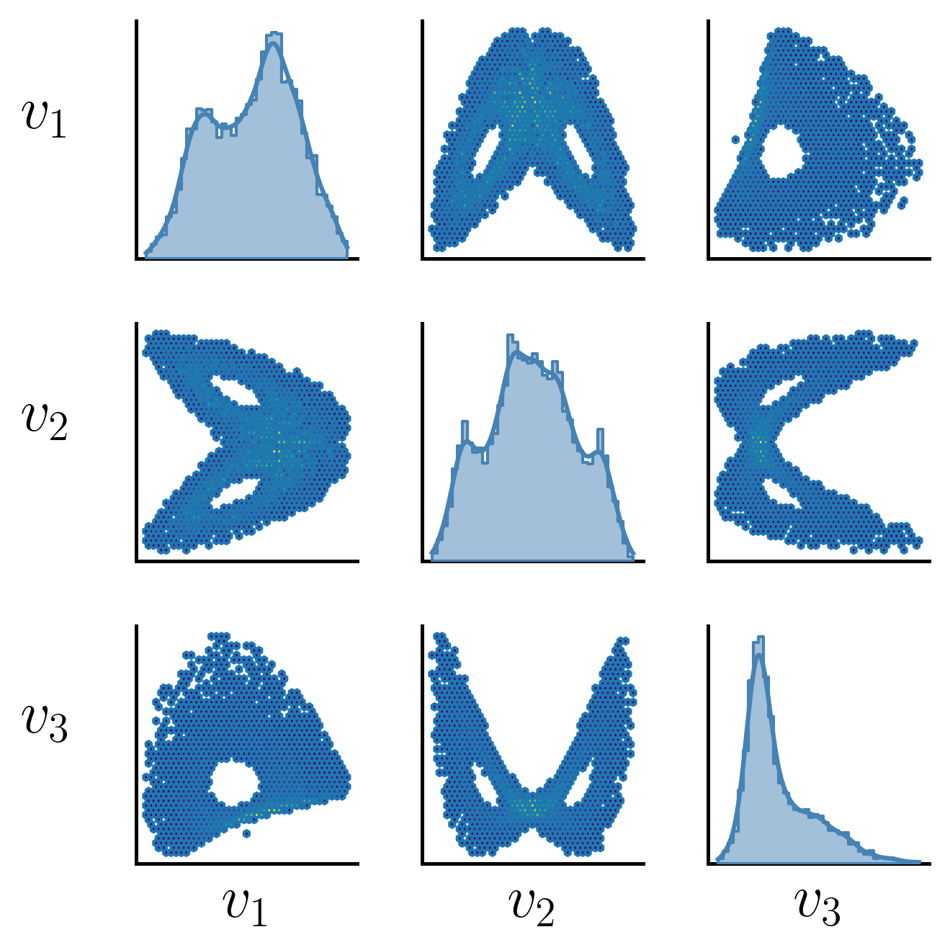}
    \caption{Generalized embedding provided by both $x_n$ and $z_n$ channels.}
    \label{fig:sub2_bn}
  \end{subfigure}
  \caption{Pairwise relationships between the principal modes in \emph{m}HAVOK embeddings. $v_i$ are the core dynamic modes.}
  \label{fig:fullwidth}
\end{figure*}

In order to demonstrate the current methodology's ability to handle several input time series, the embedding coordinates for the Lorenz system are generated for a varying number of channels, generating multiple time delay vectors (Eq.~\eqref{deylemap}). Particularly, Fig.~\ref{fig:fullwidth} shows the three most significant columns of $\mathbf{V}_c$ defined in Eq.~\eqref{VC}, i.e., the three most important linear or core dimensions, demonstrating how providing the model with an increased number of channels allows detecting complex inter-channel relationships. For instance, the pairwise relationship between $v_3$ and $v_2$ turns from an uncorrelated pattern in Fig.~\ref{fig:sub1_bn}, to a two-lobe structure reminiscent of the Lorenz attractor, visualized in Fig.~\ref{fig:sub2_bn}.

Moreover, the relationship between $v_1$ and $v_2$ changes when introducing two input channels, transitioning from a compressed geometry to the characteristic butterfly shape of the original attractor. Finally, the relationship between $v_1$ and $v_3$ appears twisted in the unidimensional embedding, as it potentially lacks vertical information from the original system.

Previous work \cite{takens_theorem_generalized, Letellier2002} already pointed out the limitations of using $f(\textbf{x}_n)=z_n$ as input channel. This is shown in Fig.~\ref{fig:sub1_bn}, where the reconstructed attractor can be seen to miss the characteristic two-lobe structure of the Lorenz system. In this work, we expand on these findings, showing how each input channel contributes unique geometrical information from the original attractor. Introducing $f(\textbf{x}_n)=x_n$ into the $f(\textbf{x}_n)=z_n$ embedding allows for the recovery of the two-lobe structure, since it includes the bimodally distributed mode $v_1$ observed in both instances of Fig.~\ref{fig:fullwidth}. However, introducing $f(\textbf{x}_n)=z_n$ alongside the $f(\textbf{x}_n)=x_n$ channel has also proven to be beneficial, as it inserts the right-skewed $v_3$ mode into the embedding, as shown in Fig.~\ref{fig:sub2_bn}. This mode appears to contain radial and vertical information from the original attractor, which is not present in the one-dimensional embedding.

Interestingly, when introducing several measurement channels, multiple independent forcing components are identified, which had to be appropriately classified and separated during \emph{m}HAVOK's regression. The methodology presented in Section~\ref{sub_sec:reg} guarantees the identification of multiple nonlinear components, filtering modes that feature $R^2$ scores lower than $\tau=0.95$ during training, as mentioned in the explanations of Eq.~\eqref{r_squared}.

\begin{table}[H]
    \renewcommand{\arraystretch}{1.5}
    \setlength{\tabcolsep}{10pt}
    \centering
    \begin{adjustbox}{scale=0.9}
    \begin{tabular}{|c|c|c|c|}
        \hline
        Channels & $r_{\text{opt}}$ & $|\mathcal{I}_c|$ & $|\mathcal{I}_f|$ \\ \hline
        $f(\mathbf{x}_n) = x_n$ & 7  & 6  & 1 \\ \hline
        $f(\mathbf{x}_n) = y_n$ & 11 & 10 & 1 \\ \hline
        $f(\mathbf{x}_n) = z_n$ & 18 & 11 & 7 \\ \hline
        $\ f(\mathbf{x}_n) = (x_n, y_n)$ & 9  & 8  & 1 \\ \hline
        $f(\mathbf{x}_n) = (x_n, z_n)$ & 8  & 6  & 1 \\ \hline
        $f(\mathbf{x}_n) = (y_n, z_n)$ & 7  & 5  & 2 \\ \hline
        $f(\mathbf{x}_n) = (y_n, z_n^2)$ & 5  & 3  & 2 \\ \hline
        $f(\mathbf{x}_n) = (x_n, y_n, z_n)$ & 18 & 15 & 3 \\ \hline
        $f(\mathbf{x}_n) = (x_n, y_n, x_n+z_n)$ & 9 & 7 & 2 \\ \hline
    \end{tabular}
    \end{adjustbox}
    \caption{Component classification of dynamical modes for different input channels using $\tau = 0.95$. 
    The optimal rank $r_{\text{opt}}$ was determined using Eq.~\eqref{eq:rank_selection_bn} with a percentile cutoff $q = 30\%$.}
    \label{tab:lorenzmulticlass}
\end{table}

Table~\ref{tab:lorenzmulticlass} confirms the presence of several components with a low $R^2$ score during \emph{m}HAVOK's component classification. Additionally, it shows how single channel embeddings are susceptible to increased noise correlation, as illustrated by the component classification of $f(\textbf{x}_n)=z_n$, where the number of nonlinear components is significantly higher than for other single channel embeddings. Hence, this particular embedding potentially contains noisy modes. On the other hand, the generalized $f(\textbf{x}_n)=(x_n,z_n)$ embedding features only one nonlinear component, demonstrating how an extended embedding may contribute to well-behaved \emph{m}HAVOK reconstructions. 
%which solemnly feature true forcing components. 

As noticed, $r_{\text{opt}}$ changes depending on the input channels. For instance, attempting a single channel reconstruction for $z_n$ using $r=11$ results in a disorganized reconstructed attractor, while it guarantees a successful reconstruction for $y_n$. Moreover, selecting appropriate measurement time series can drastically reduce the optimal rank value. For example, performing linear combinations of input channels may reduce the number of required dynamical modes, likely because of uncorrelated noise components from each channel canceling out when combined.

%\newpage

\begin{figure*}[htbp]
  \centering
  \begin{subfigure}{0.31\textwidth}
    \includegraphics[width=\linewidth]{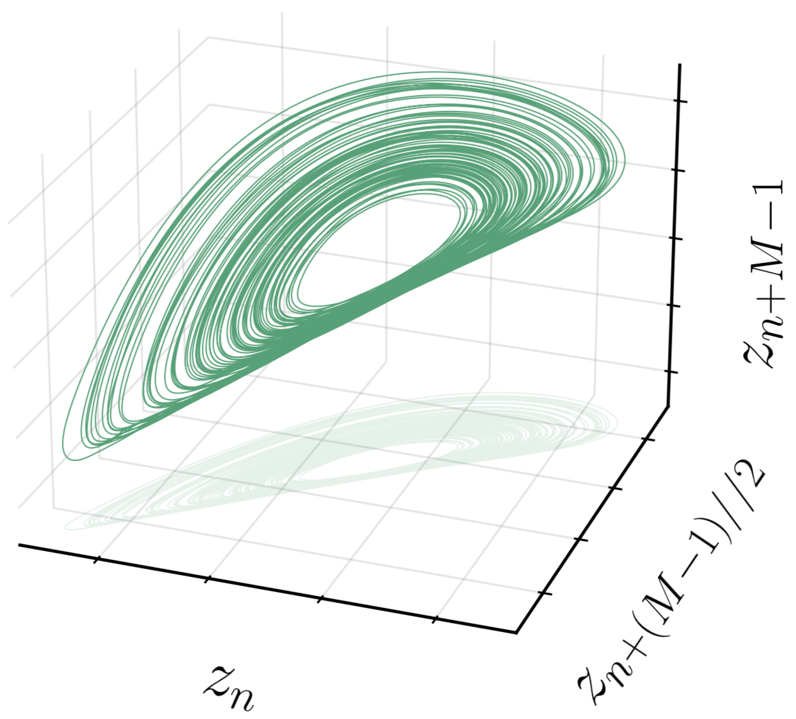}
    \caption{Single channel reconstruction using $f(\textbf{x}_n)=(z_n)$ as input.}
    \label{fig:sub1summ}
  \end{subfigure}
  \hfill
  \begin{subfigure}{0.31\textwidth}
    \includegraphics[width=\linewidth]{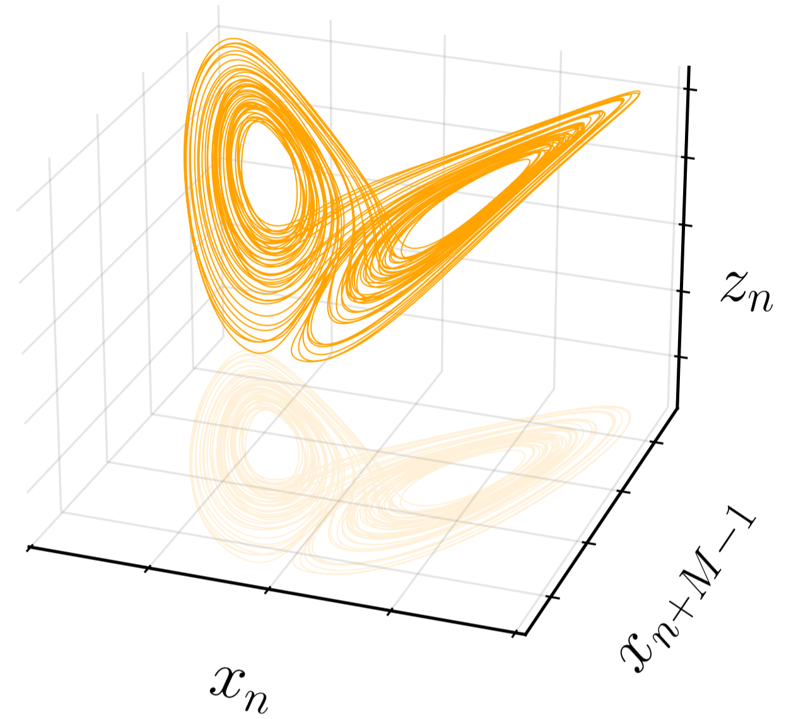}
    \caption{Multichannel reconstruction using $f(\textbf{x}_n)=(x_n,z_n)$ as input.}
    \label{fig:sub2summ}
  \end{subfigure}
  \hfill
  \begin{subfigure}{0.31\textwidth}
    \includegraphics[width=\linewidth]{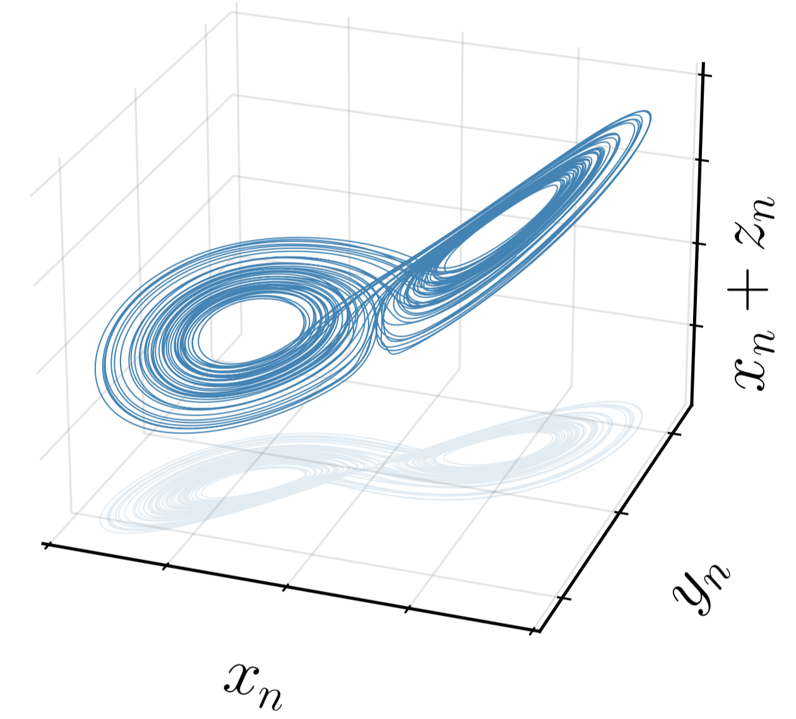}
    \caption{Multichannel reconstruction using $f(\textbf{x}_n)=(x_n,y_n,x_n+z_n)$ as input.}
    \label{fig:sub3summ}
  \end{subfigure}
  \caption{Reconstructed attractors in the original coordinates for multiple observables.}
  \label{fig:summaryfig}
\end{figure*}

% \begin{figure}[H]
%     \centering
%     \begin{subfigure}{0.46\linewidth}
%         \includegraphics[width=\linewidth]{Images/1_Lorenz/barplot_x.png}
%         \caption{Input channel: $x_n$.}
%     \end{subfigure}
%     \hfill
%     \begin{subfigure}{0.46\linewidth}
%         \includegraphics[width=\linewidth]{Images/1_Lorenz/barplot_z.png}
%         \caption{Input channel: $z_n$.}
%     \end{subfigure}

% \end{figure}
   
% \begin{figure}[H]\ContinuedFloat
%     \centering
%     \begin{subfigure}{0.48\linewidth}
%         \includegraphics[width=\linewidth]{Images/1_Lorenz/barplot_yz.png}
%         \caption{Input channels: $(y_n, z_n)$.}
%     \end{subfigure}
%     \hfill
%     \begin{subfigure}{0.48\linewidth}
%         \includegraphics[width=\linewidth]{Images/1_Lorenz/barplot_xyz.png}
%         \caption{Input channels: $(x_n,y_n,z_n)$.}
%     \end{subfigure}
%         \caption{Component classification for different input channels. 
%         The dotted line indicates $\tau = 0.95$. 
%         Green components are linear, red are nonlinear.}
%     \label{fig:BARRAS}
% \end{figure}

In order to assess the success of the algorithm in each of the presented paradigms, each of the reconstructed time series is transformed back to the original coordinate system using the methodology in Section~\ref{subsec:reconstructed_attractor}. Fig.~\ref{fig:summaryfig} displays the reconstructed attractors for some input channel embeddings in the original coordinate space. First off, Fig.~\ref{fig:sub1summ} showcases how the embedding of $f(\textbf{x}_n)=z_n$ fails to represent the original attractor's topology, as reported in References \cite{Letellier2002, takens_theorem_generalized}. Then, Fig.~\ref{fig:sub2summ} shows how \emph{m}HAVOK successfully reproduces the attractor presented in Deyle and Sugihara's work, where two input channels $f(\textbf{x}_n)=(x_n,z_n)$ are included \cite{takens_theorem_generalized}. Complementarily, the algorithm's ability to reproduce the Lorenz attractor is showcased in Fig.~\ref{fig:sub3summ}, where three input channels $f(\textbf{x}_n)=(x_n,y_n,x_n+z_n)$ are provided. Interestingly, the reconstructed attractor is visually similar to the original, even when a linear combination of state variables is included in one of the input channels. Such a result is useful when reconstructing a broader range of real world dynamical systems from experimental data, where multiple sensor readings are available. In these cases, it is common that complicated functions of the system's state variables are measured rather than the state variables themselves. 

\begin{figure}[H]
    \centering
    \begin{subfigure}{0.46\linewidth}
        \includegraphics[width=\linewidth]{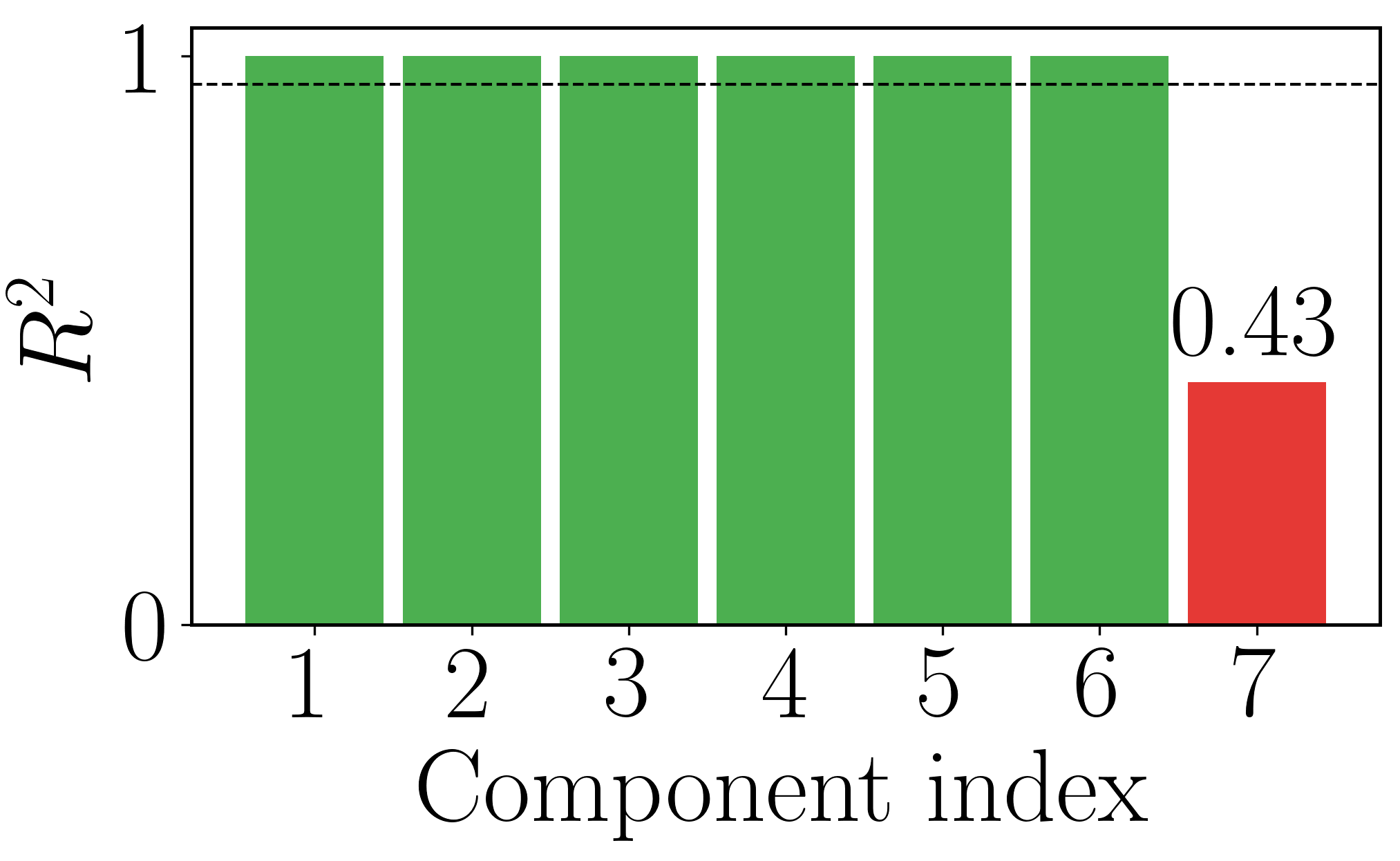}
        \caption{Input channel: $x_n$.}
    \end{subfigure}
    \hspace{1em}
    \begin{subfigure}{0.46\linewidth}
        \includegraphics[width=\linewidth]{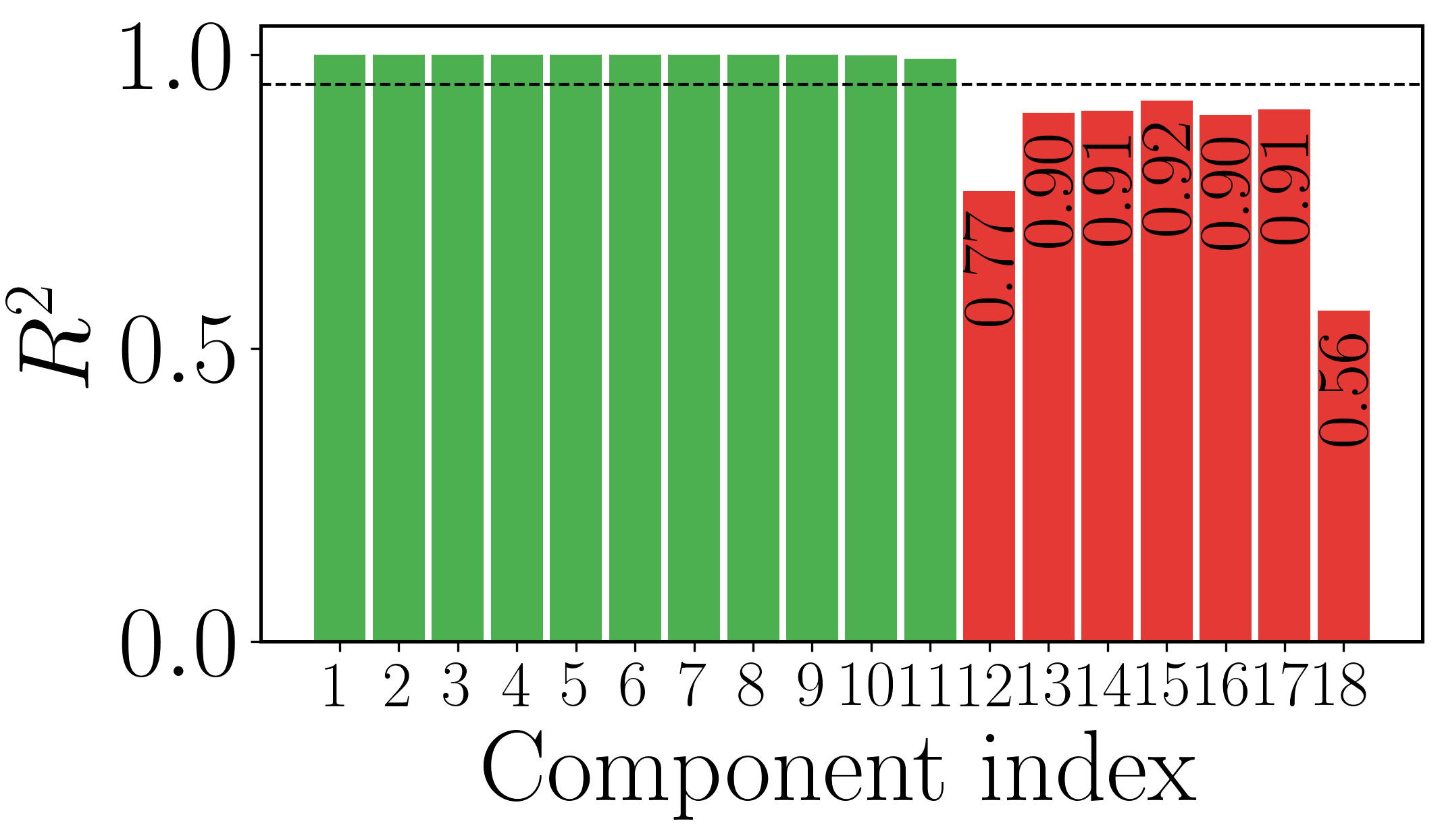}
        \caption{Input channel: $z_n$.}
    \end{subfigure}

    \vspace{1em} % space between rows

    \begin{subfigure}{0.46\linewidth}
        \includegraphics[width=\linewidth]{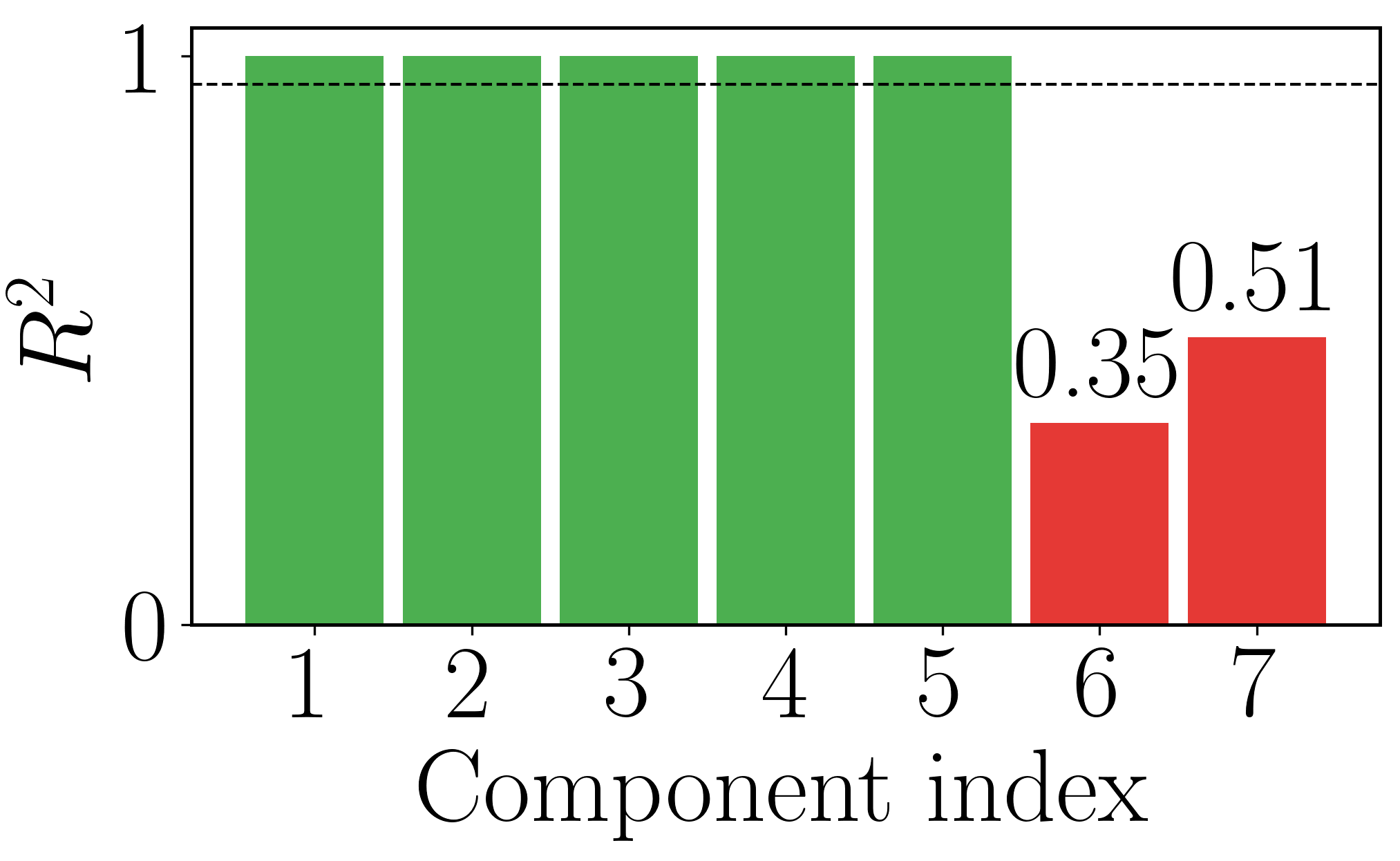}
        \caption{Input channels:\\$(y_n, z_n)$.}
    \end{subfigure}
    \hspace{1em}
    \begin{subfigure}{0.46\linewidth}
        \includegraphics[width=\linewidth]{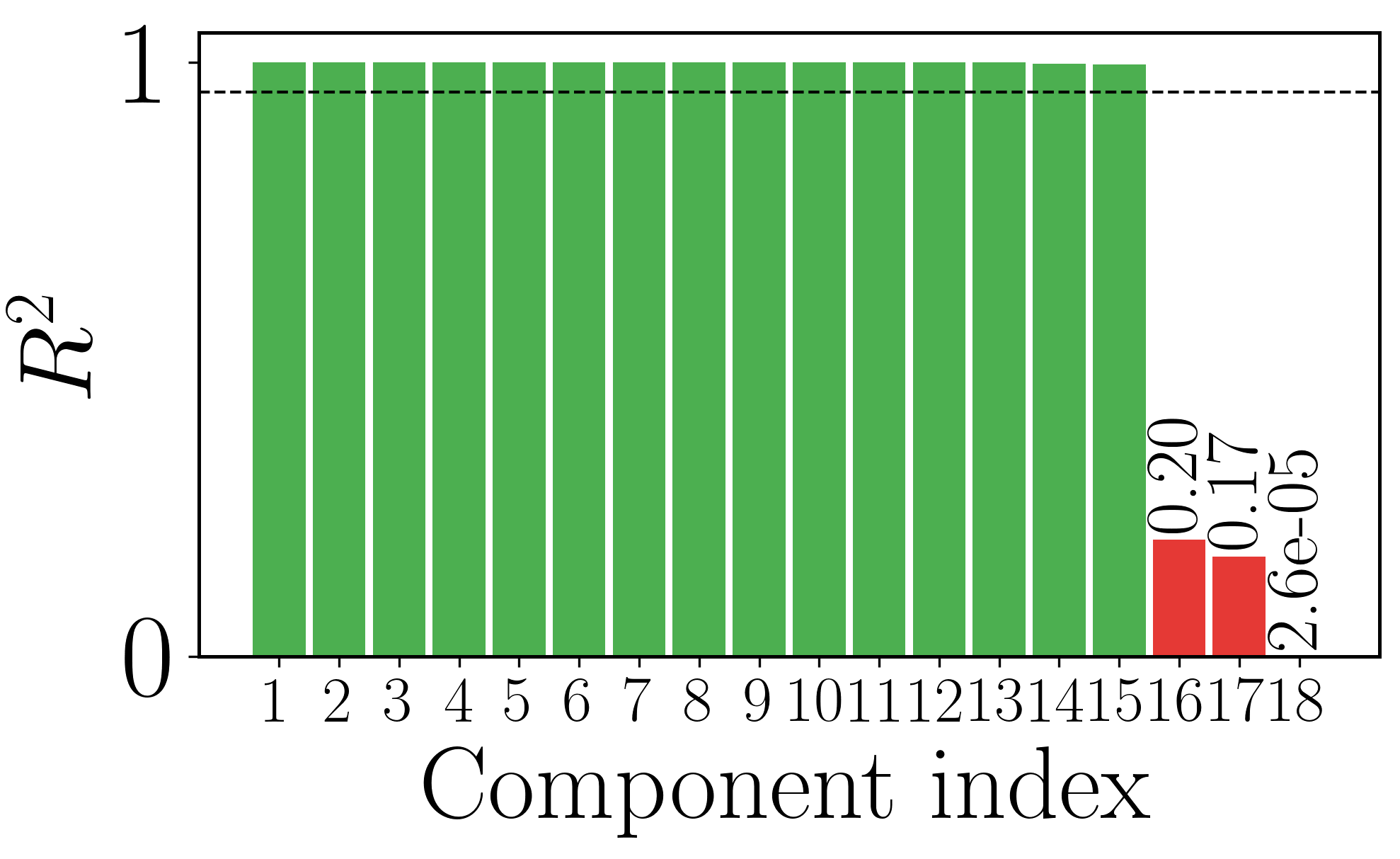}
        \caption{ \centering Input channels:\\$(x_n,y_n,z_n)$.}
    \end{subfigure}

    \caption{Component classification for different input channels. 
    The dotted line indicates $\tau = 0.95$. 
    Green components are linear, red are nonlinear.}
    \label{fig:BARRAS}
\end{figure}

Fig.~\ref{fig:BARRAS} exhibits the $R^2$ score defined in Eq.~\eqref{r_squared} for multiple input channels. In consistency with the findings of Brunton et al. \cite{Brunton_Havok_Nature}, the last component in the embedding has a low $R^2$ score in comparison to those from its neighbors. As shown in Fig.~\ref{fig:BARRAS}, multiple channel embeddings may contain more than one nonlinear component, generalizing Brunton's assertion. 

\begin{figure}[H]
    \centering
  
    \begin{subfigure}{0.45\linewidth}
        \centering
        \includegraphics[width=\linewidth]{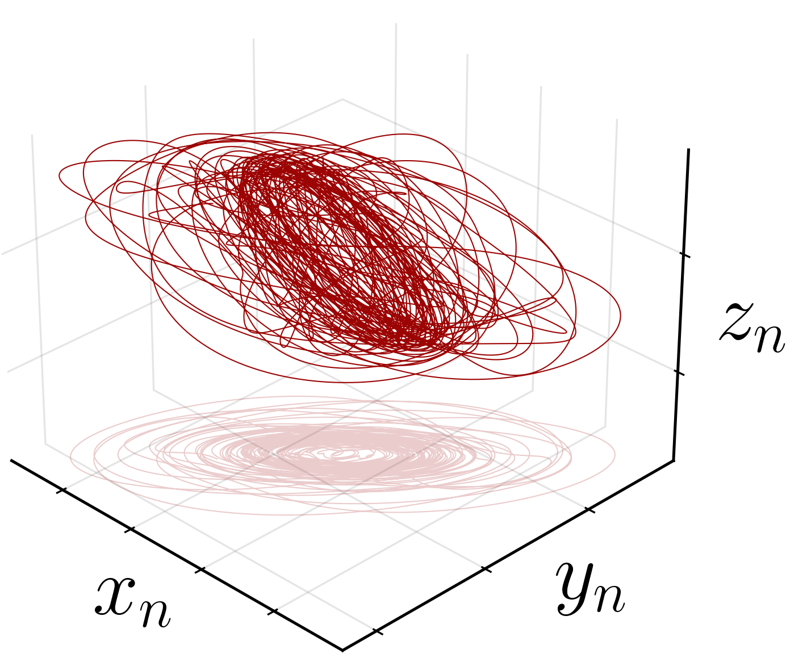}
        \caption{}
        \label{fig:havok_class}
    \end{subfigure}
    %\hfill
    \begin{subfigure}{0.45\linewidth}
        \centering
        \includegraphics[width=\linewidth]{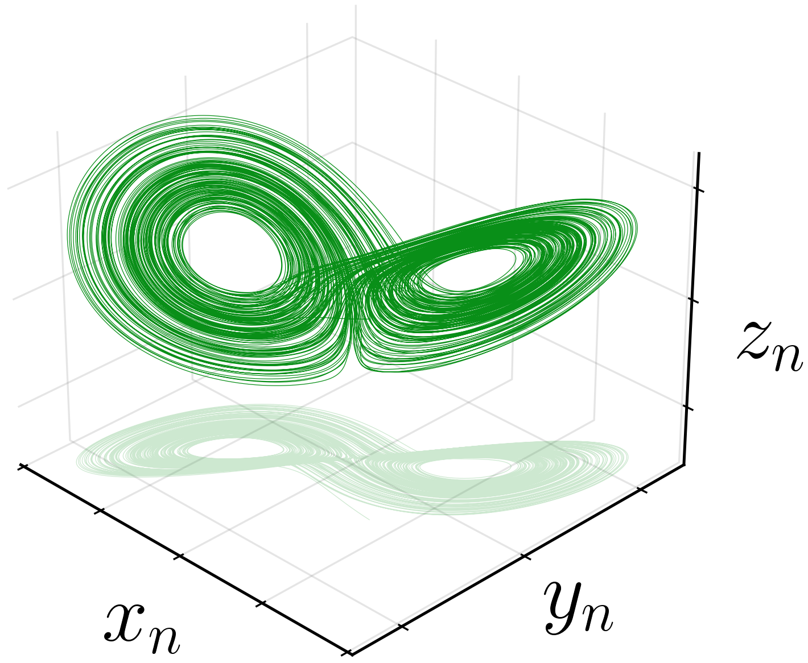}
        \caption{}
        \label{fig:mhavok_class}
    \end{subfigure}
    
    \caption{Reconstruction of the Lorenz attractor using (a) the component classification analogous to the original HAVOK method, i.e., considering only one forcing component, and (b) using the component classification scheme proposed by \emph{m}HAVOK. The input channels were $(x_n,y_n,z_n)$.}
    \label{fig:comparison_component_classification}
\end{figure}

Additionally, Fig.~\ref{fig:comparison_component_classification} confirms how this generalization is an essential step to reconstruct the system's dynamics from multiple observables. If only the last dynamical mode is used for forcing, the system's reconstruction remarkably degrades, as demonstrated by the marked difference between Fig.~\ref{fig:havok_class}, where only the last component has been considered as the forcing term, and Fig.~\ref{fig:mhavok_class}, where all nonlinear components were used for forcing. 
%Furthermore, it is important to notice how the current methodology is capable of identifying bad-performing modes that are not necessarily located last in the classification.  

\section{Results - Sprott system} \label{sec:results_sprott}

In comparison to the Lorenz attractor, the reconstruction of Sprott system poses a major challenge, since depending on the initial conditions, the system evolves to a basin of attraction or to an invariant set, as mentioned in Section~\ref{subsec:test_systems_studied}. For instance, the invariant torus shown in Fig.~\ref{fig:sprotts_original_attractor} is neutrally stable for trajectories initialized precisely on it, but does not attract an open neighborhood in phase space \cite{SPROTT20141361}. Consequently, marginally distinct initial conditions will send the trajectory outside the torus and towards the strange attractor.

\subsection{Generalized embeddings}\label{sec:sprott_generalized_embeddings}

As previously mentioned, the Sprott system features two types of solutions: the conservative torus and the strange attractor, which resembles a cord. Interestingly, this system is invariant under the transformation $\mathcal{G}:(x,y,z)\xrightarrow{}(x,-y,-z)$ \cite{SPROTT20141361}. The torus solutions feature a $180^\circ$ rotational symmetry about the $x$-axis, while the cord is an attractor in forward time and a repellor in reversed time, as well as symmetric under a $180^\circ$ rotation about the $x$-axis. Recalling the argument by Letellier and Aguirre, symmetry-blind observables defined in Section~\ref{subsec:generalized_embedding} will fail to reconstruct the attractors' topology \cite{Letellier2002}. 

\begin{figure}[H]
  \centering
  % Row 1
  \begin{subfigure}{0.23\textwidth}
    \includegraphics[width=\linewidth]{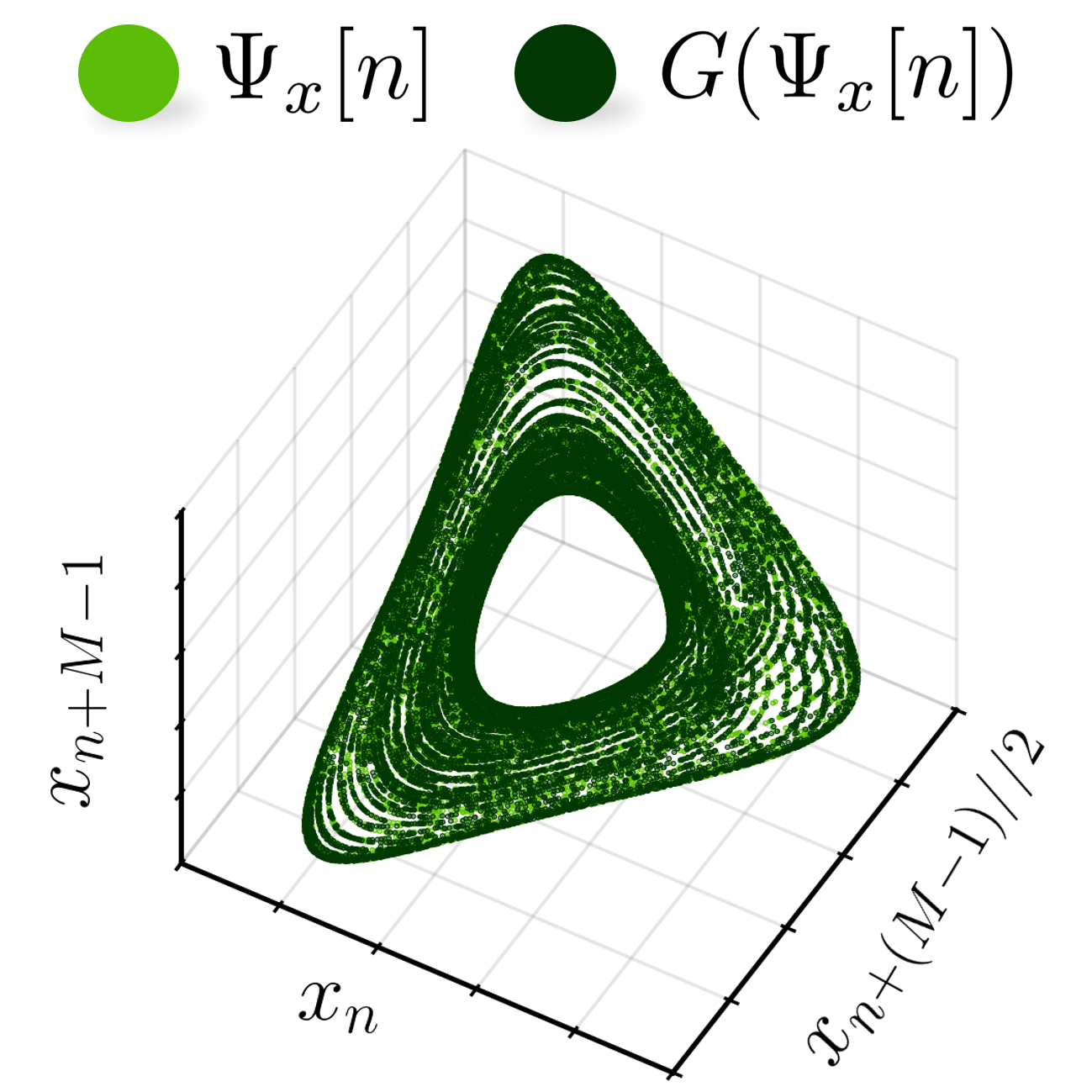}
    \caption{}
    \label{fig:single2}
  \end{subfigure}
  %\hfill
  \begin{subfigure}{0.23\textwidth}
    \includegraphics[width=\linewidth]{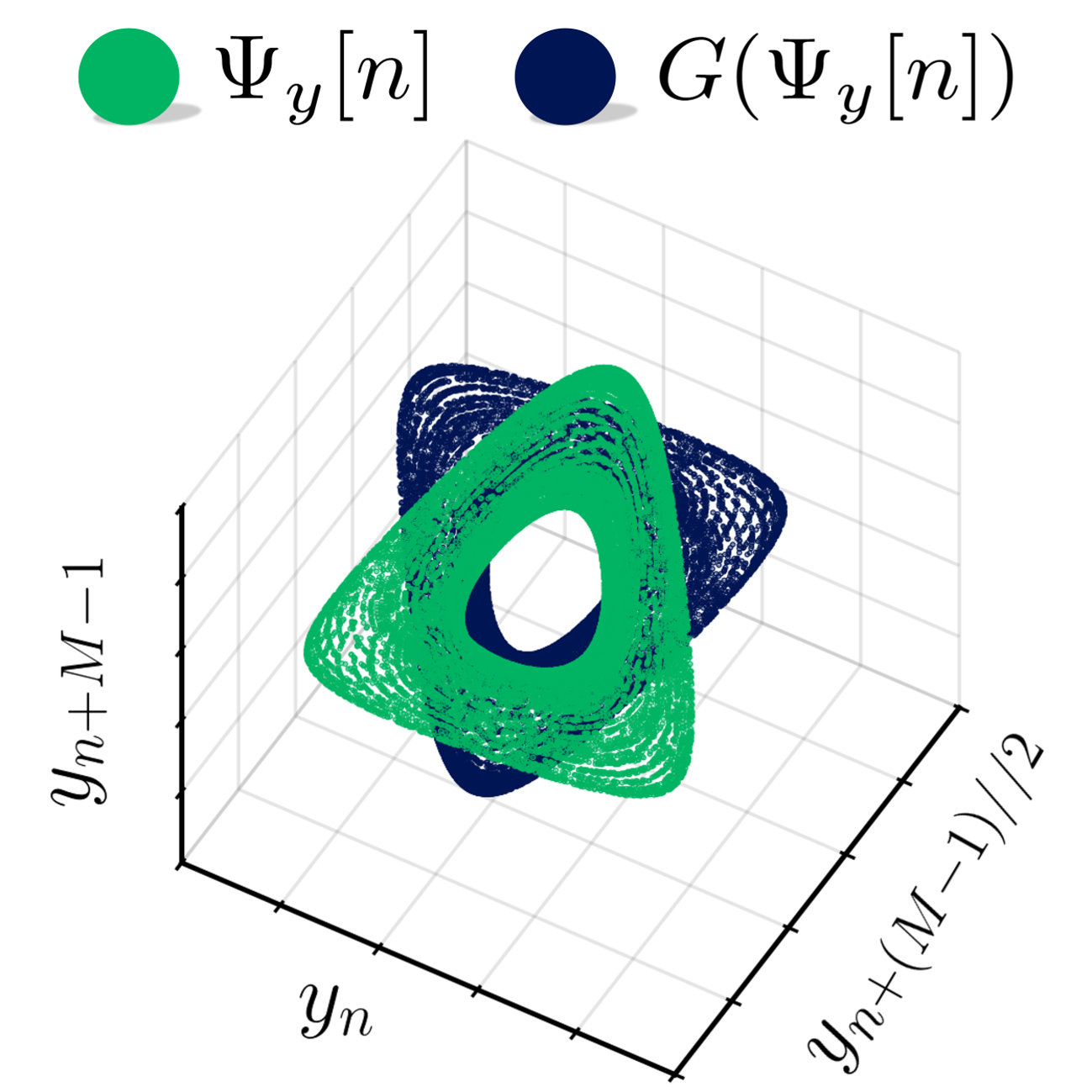}
    \caption{}
    \label{fig:multi22}
  \end{subfigure}

  \vskip\baselineskip % space between rows

  % Row 2
  \begin{subfigure}{0.23\textwidth}
    \includegraphics[width=\linewidth]{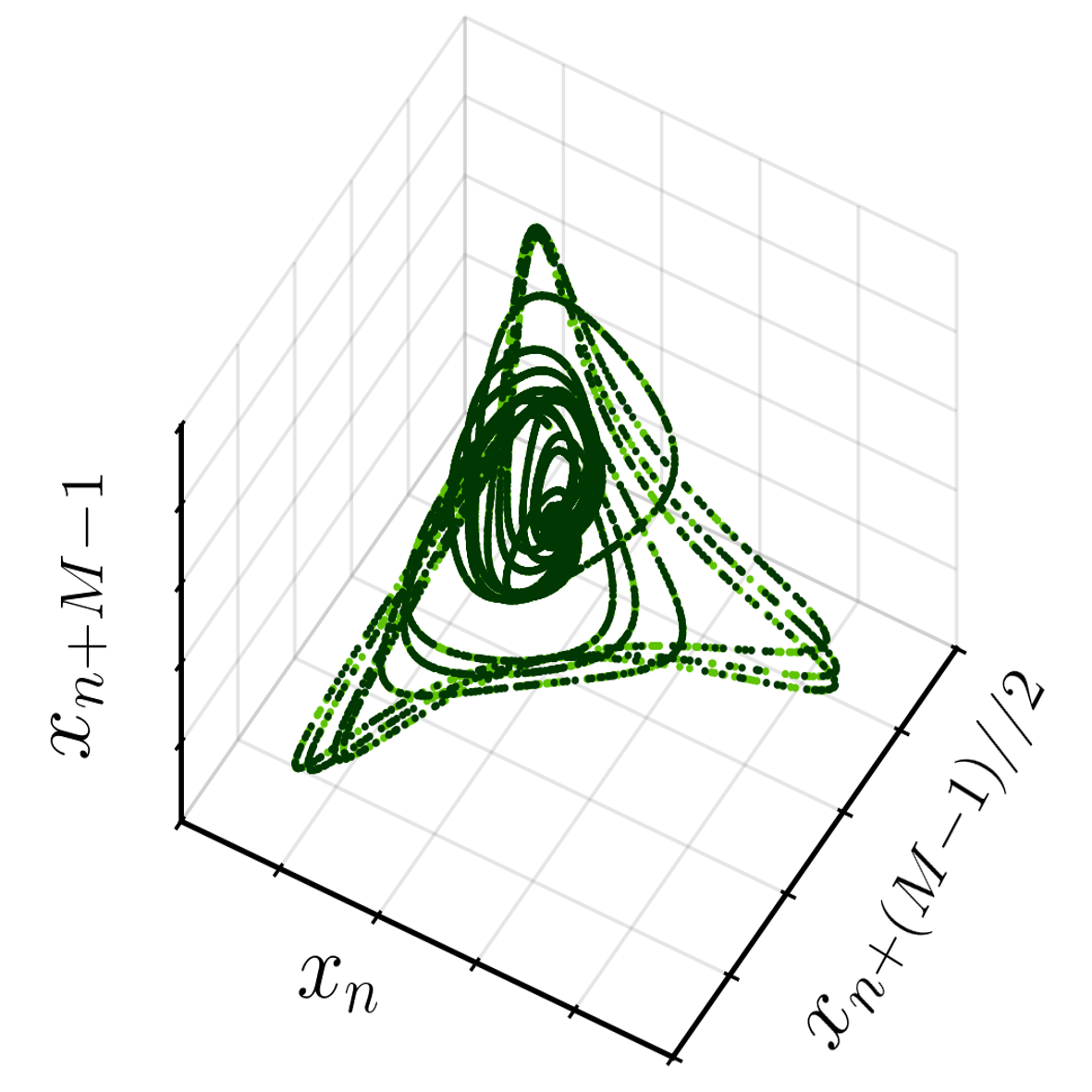}
    \caption{}
    \label{fig:multi23}
  \end{subfigure}
  %\hfill
  \begin{subfigure}{0.23\textwidth}
    \includegraphics[width=\linewidth]{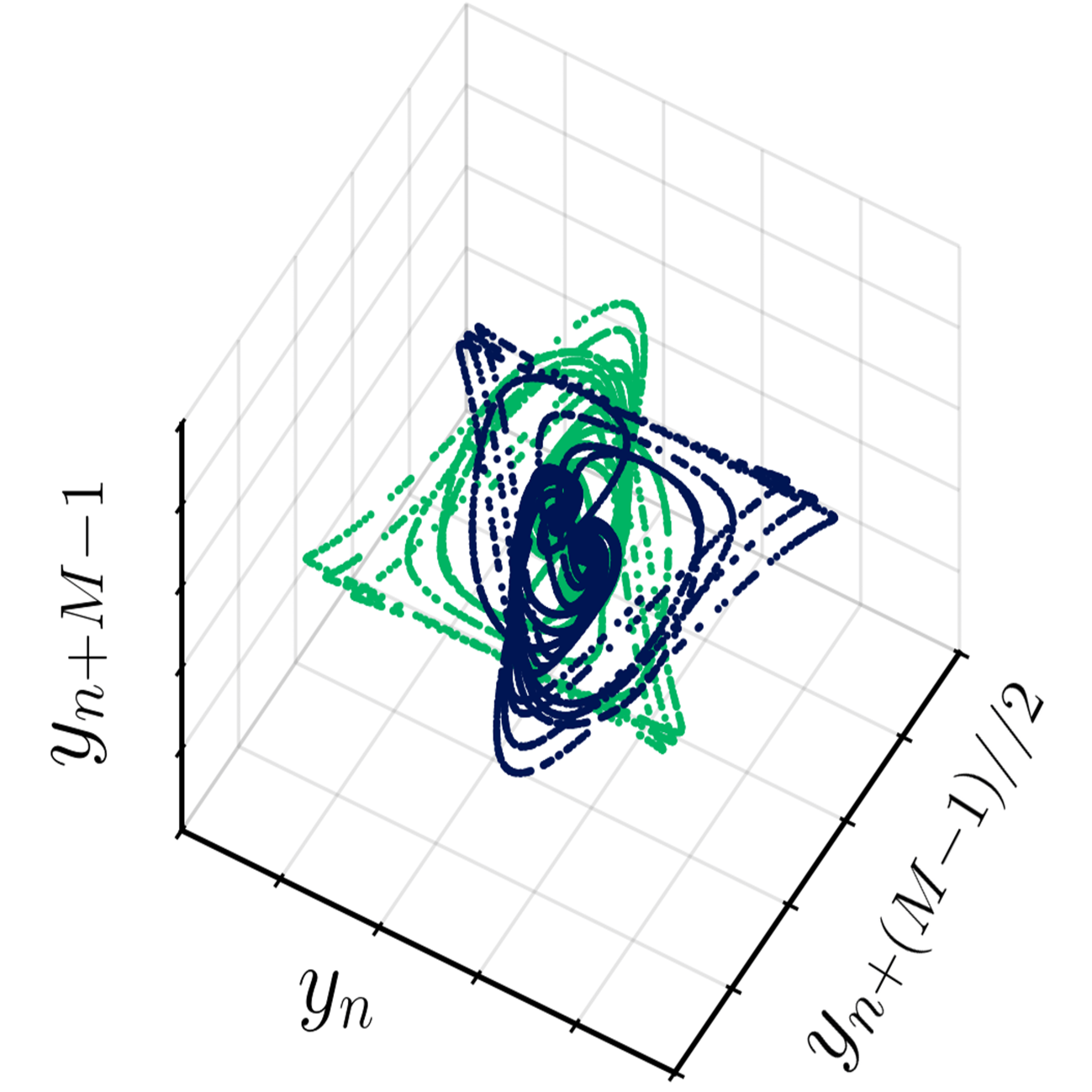}
    \caption{}
    \label{fig:multi24}
  \end{subfigure}

  \caption{Reconstructions in real space of the Sprott system based on single-channel delay embeddings and their image under $\mathcal{G}$. $(a,b)$ Trajectories located on the invariant torus with input channels $x_n$ and $y_n$. $(c,d)$ Trajectories near or on the chord-like attractor with input channels $x_n$ and $y_n$.}
  \label{fig:delays_G}
\end{figure}

Unlike the Lorenz attractor, Sprott system is highly susceptible to the choice of observables and often requires several measurements of the state variables to recover either the strange attractor, the invariant torus, or both. A single time series delay map $\Psi[n]$ was constructed using information from $f(\textbf{x}_n)=x_n$ and $f(\textbf{x}_n)=y_n$ with $M=1000$. Then, the transformation $\mathcal{G}$ was applied to verify invariance of observables. Fig.~\ref{fig:delays_G} demonstrates the symmetry blindness of $\mathcal{G}$-even measurements from both the torus and the strange attractor. As can be noticed, even observables such as $x_n$ coincide with their $\mathcal{G}$ images. In contrast, odd observables are not set-wise coincident. Hence, the attractor features $\mathcal{G}$-even symmetry blindness.

%% CUIDADO CON LOS LABELS X,Y,Z. ESTAMOS EN EL ESPACIO DE TAKENS, ES DECIR SON VECTORS THE TAKENS PSI_X[N], PSI_X[N+500], PSI_X[N+1000]
% \begin{figure}[H]
%   \centering
%   \begin{subfigure}{0.25\textwidth}
%     \includegraphics[width=\linewidth]{Images/2_Sprott/x-embedding_symm.png}
%     \caption{Input channel: $x_n$.}
%     \label{fig:single2}
%   \end{subfigure}%\hfill
%   \begin{subfigure}{0.25\textwidth}
%     \includegraphics[width=\linewidth]{Images/2_Sprott/y-embedding_symm.png}
%     \caption{Input channel: $y_n$.}
%     \label{fig:multi22}
%   \end{subfigure}
%   \caption{Single channel delays of the torus and their image under $\mathcal{G}$.} \label{fig:torus_embeddings}
% \end{figure}
The apparent symmetry blindness in both the torus and the strange attractor relates directly to the previously mentioned findings in Ref.~\cite{Letellier2002}, where it is shown that observables invariant under a system's symmetry do not provide any information about the lost equivariance, obscuring reconstructions. The authors classified these reconstructions as image systems, where attractors might be dynamically equivalent but stripped of the underlying symmetry properties \cite{Letellier2002}.

\begin{figure*}[htbp]
  \centering
  \begin{minipage}{0.25\textwidth}
    \centering
    \begin{subfigure}{\linewidth}
      \centering
      \includegraphics[width=\linewidth]{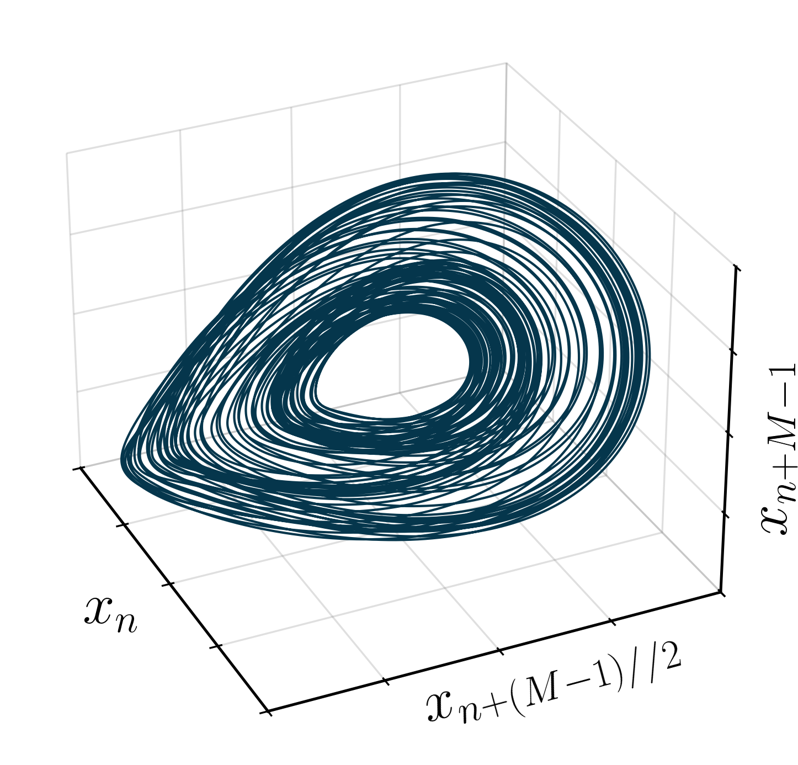}
      \caption{Symmetry-blind reconstruction for the torus using $f(\mathbf{x}_n)=x_n$ as input.}
      \label{fig:sprott_x1}
    \end{subfigure}
    \vskip\baselineskip
    \begin{subfigure}{\linewidth}
      \centering
      \includegraphics[width=\linewidth]{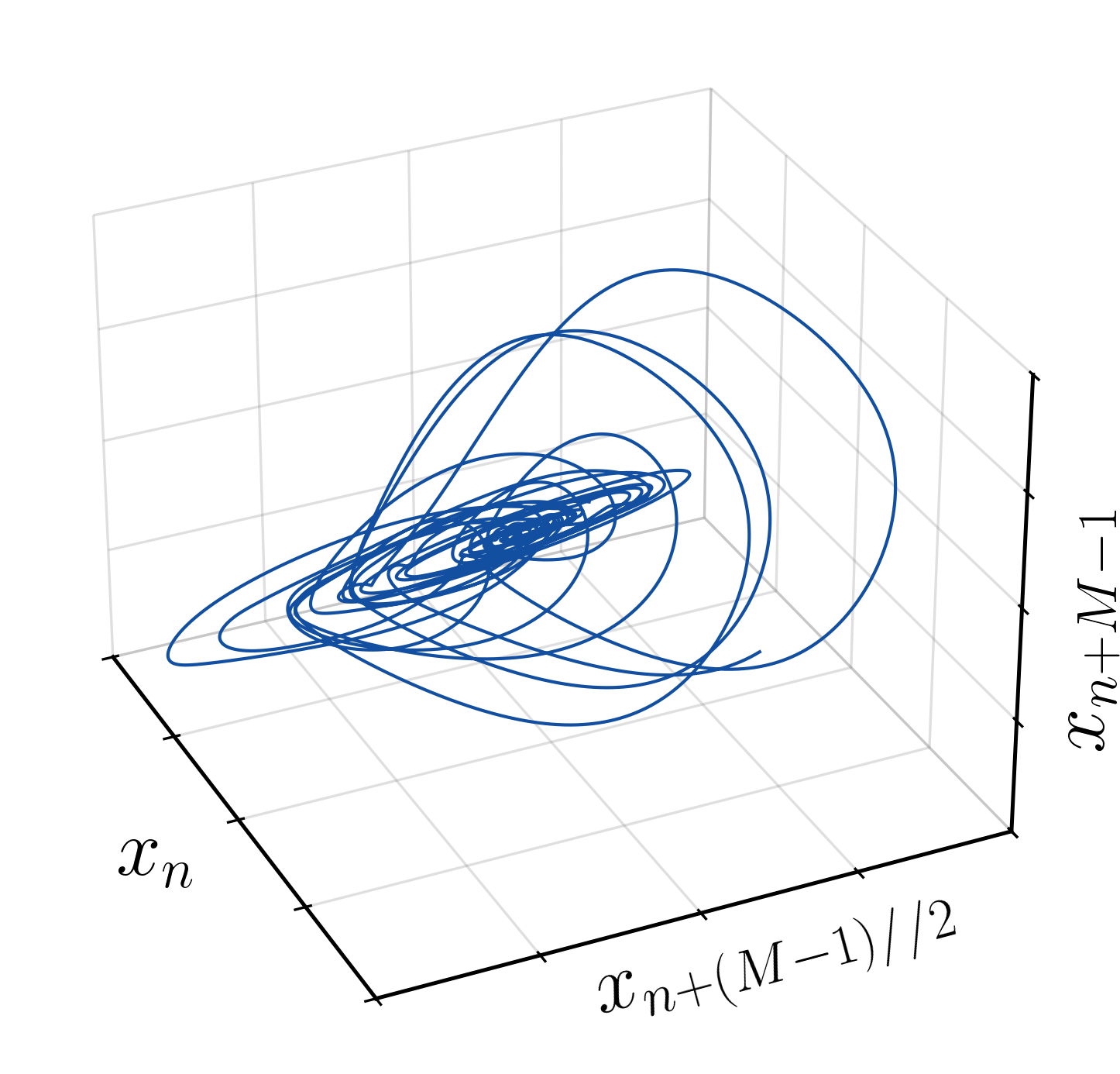}
      \caption{Symmetry-blind reconstruction for the strange attractor using $f(\mathbf{x}_n)=x_n$ as input.}
      \label{fig:sprott_y1}
    \end{subfigure}
  \end{minipage}
  \hspace{1em}
  \begin{minipage}{0.25\textwidth}
    \centering
    \begin{subfigure}{\linewidth}
      \centering
      \includegraphics[width=\linewidth]{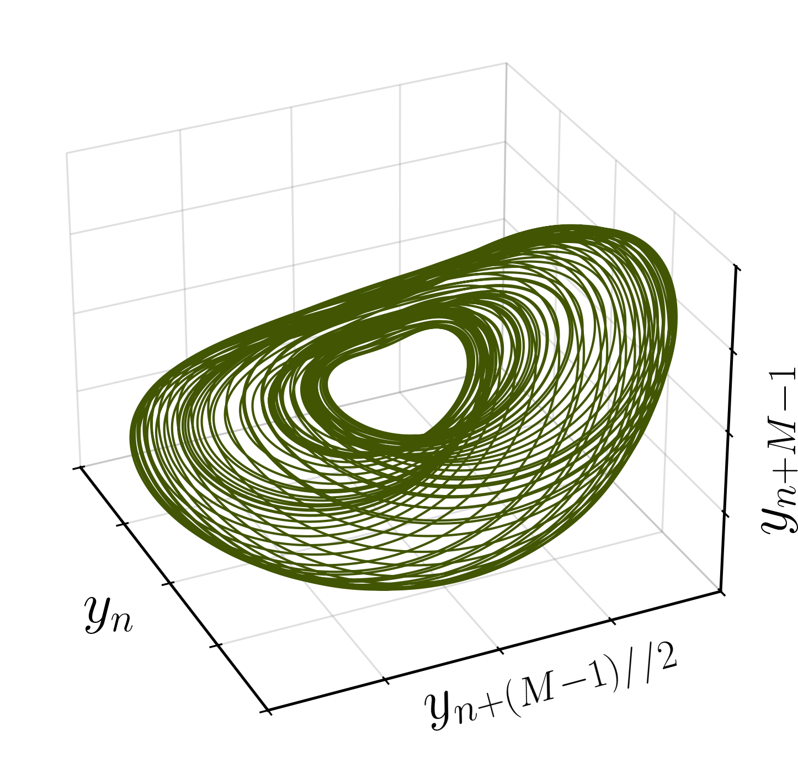}
      \caption{Torus reconstruction using $f(\mathbf{x}_n)=y_n$ as a single input.}
      \label{fig:sprott_x2}
    \end{subfigure}
    \vskip\baselineskip
    \begin{subfigure}{\linewidth}
      \centering
      \includegraphics[width=\linewidth]{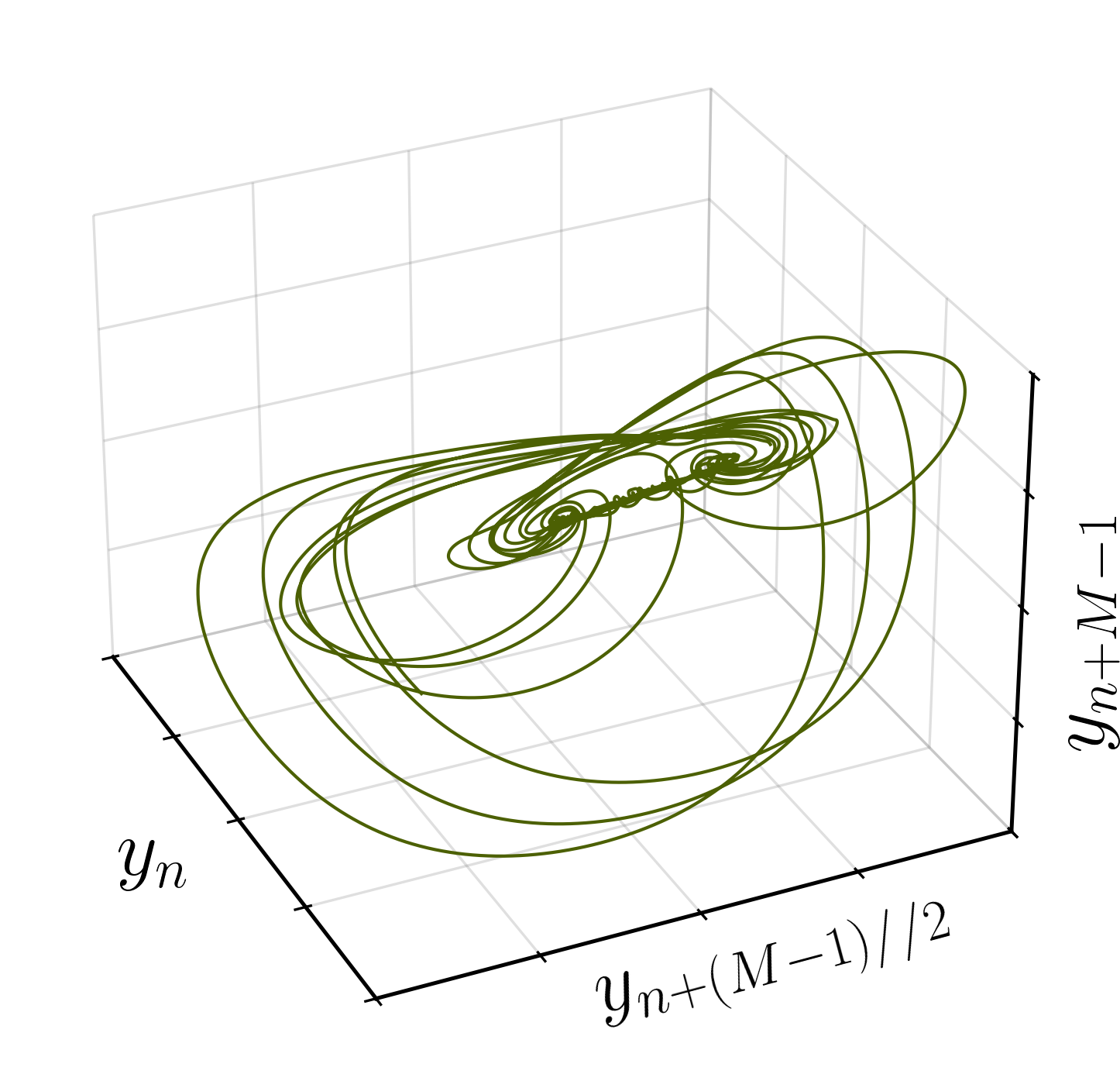}
      \caption{Chord reconstruction using $f(\mathbf{x}_n)=y_n$ as a single input.}
      \label{fig:sprott_y2}
    \end{subfigure}
  \end{minipage}
  \hspace{1em}
  \begin{minipage}{0.25\textwidth}
    \centering
    %\vspace{0.01\textheight} % pushes the figure down toward the vertical middle
    \begin{subfigure}{\linewidth}
      \centering
      \includegraphics[width=\linewidth]{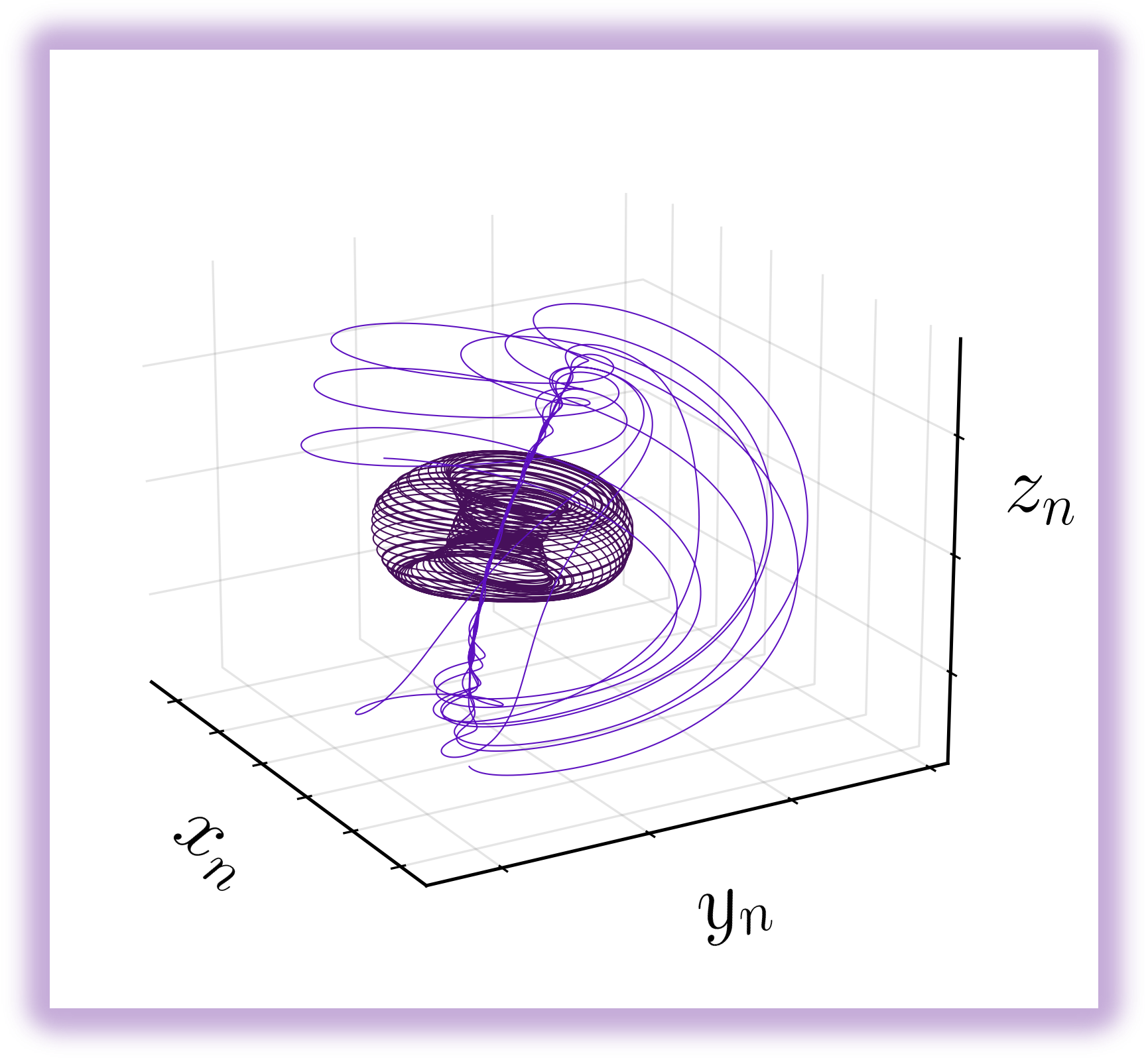}
      \caption{Reconstruction using $(x_n,y_n,z_n)$ from both initial conditions.}
      \label{fig:sprott_z}
    \end{subfigure}
  \end{minipage}

  \caption{Reconstructed attractors in the original coordinates for different channel inputs.}
  \label{fig:sprott_gallery}
\end{figure*}

Due to the symmetry blindness of the torus' observables, state-space reconstruction using \emph{m}HAVOK may require multiple observables. As previously mentioned, generalized embeddings provide the necessary machinery by combining several measurement channels, even across different initial conditions. In Fig.~\ref{fig:sprott_gallery}, \emph{m}HAVOK was run with different combinations of input channels, plotting the reconstructed attractors employing the methodology in Section~\ref{subsec:reconstructed_attractor}. As shown, single even channel reconstructions collapse the torus in Fig.~\ref{fig:sprott_x1}, obscuring the system's symmetry properties. When single odd channel reconstructions were performed, the reconstructions better mirror the original system, as shown in Figs.~\ref{fig:sprott_x2} and ~\ref{fig:sprott_y2}. These results confirm the fact that symmetry blind observables negatively impact \emph{m}HAVOK's ability to reconstruct attractors by obscuring geometrical features; the multichannel generalization provides a safe mechanism when dealing with symmetry blindness, as demonstrated on Fig.~\ref{fig:sprott_z}, where all three channels were used in the reconstruction.

The dissipation $\langle y+z\rangle$ defined in \cite{SPROTT20141361} was calculated for the reconstructed observables in Fig.~\ref{fig:sprott_z}. The dissipation for the torus' reconstruction was $\langle y_1+z_1\rangle=0.001$, in agreement with Sprott's work \cite{SPROTT20141361}. On the other hand, the dissipation for the attractor's reconstruction was, $\langle y_2+z_2\rangle=-0.044$, indicating negative dissipation within the basin of attraction, as indicated by Ref. \cite{SPROTT20141361}. These results demonstrate that delay maps can faithfully reproduce geometric and dynamical properties given sufficient observable information. 

\begin{figure*}[htbp]
    \centering
    \includegraphics[width=0.82\linewidth]{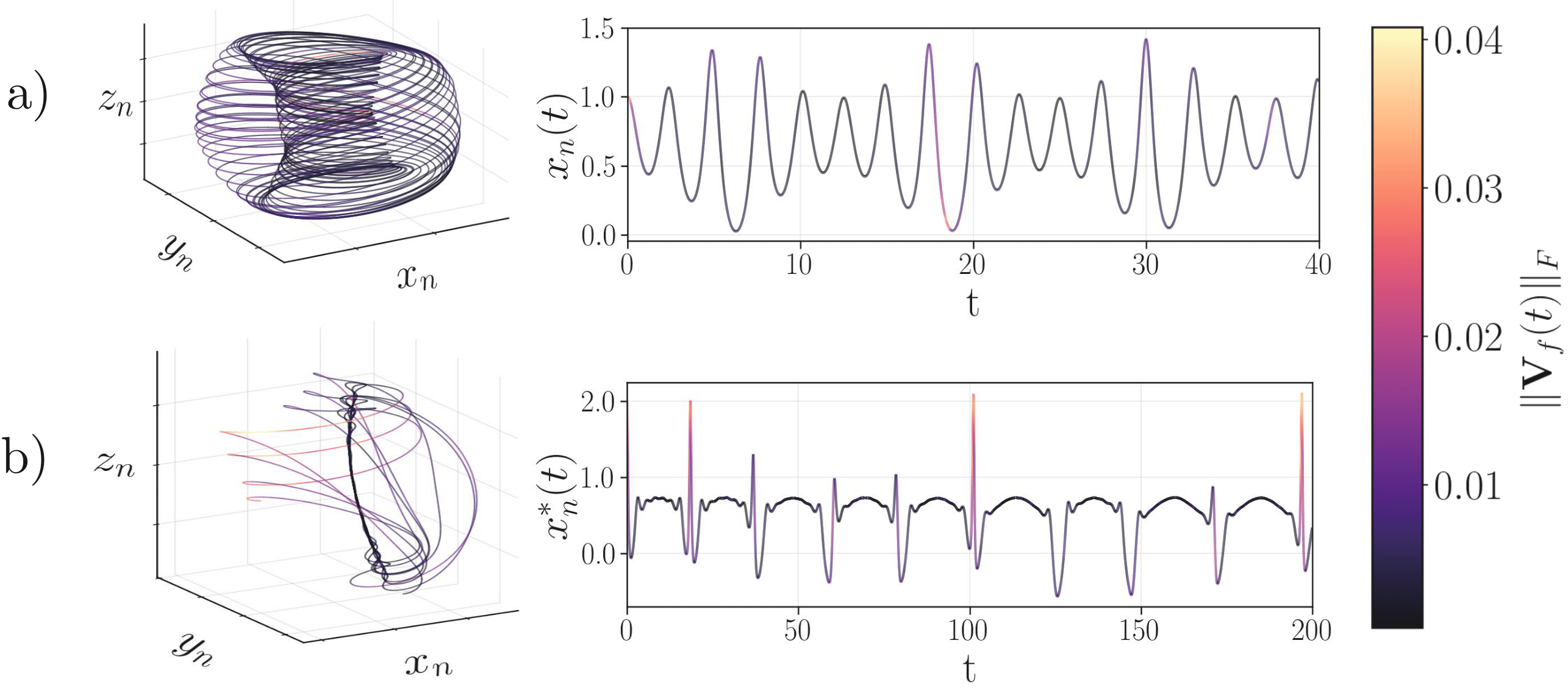}
    \caption{Nonlinear components $\textbf{V}_f$ identified by \emph{m}HAVOK for a) the torus and b) the strange attractor.}
    \label{fig:sprott_forcing}
\end{figure*}

Beyond geometry and dissipation, it is important to evaluate the reconstruction's ability to classify nonlinear components that drive the dynamics. Similar to the work by Brunton et al. \cite{Brunton_Havok_Nature}, \emph{m}HAVOK was employed as a diagnostic tool to identify nonlinear forcing contributions, as visualized in Fig.~\ref{fig:sprott_forcing}.

Reportedly, the conservative torus has Lyapunov exponents $(0,0,0)$ \cite{SPROTT20141361}. Therefore, \emph{m}HAVOK should in principle return a purely linear model, which is largely confirmed by the time series of Fig.~\ref{fig:sprott_forcing}\textcolor{blue}{a}. In practice, the appearance of small nonlinear terms in the torus is attributed to numerical noise and truncation artifacts, highlighting the need for a posterior analysis tool that discerns noise from true forcing. On the other hand, Fig.~\ref{fig:sprott_forcing}\textcolor{blue}{b} features nonlinear terms during intermittent events. Therefore, they are likely to be related to true system forcing, making them essential for a faithful reconstruction.

\subsection{Automated rank selection algorithm}\label{sub_sec:rank}

Throughout this section, \emph{m}HAVOK was run by simultaneously using the channels $f(\textbf{x}_n)=(x_n,y_n,z_n)$ from both initial conditions, which are necessary to reconstruct the invariant torus and the chord-like attractor, respectively. The rationale behind this decision will be discussed in the following (Subsection~\ref{sec:sprott_quantitative_evaluation}). 

\subsubsection{\texorpdfstring{Variance-informed $r$ selection}{Variance-informed r selection}}
\label{sub_sec:results_variance_rank_selection_attempt}

\begin{figure}[H]
    \centering
    \includegraphics[width=\linewidth]{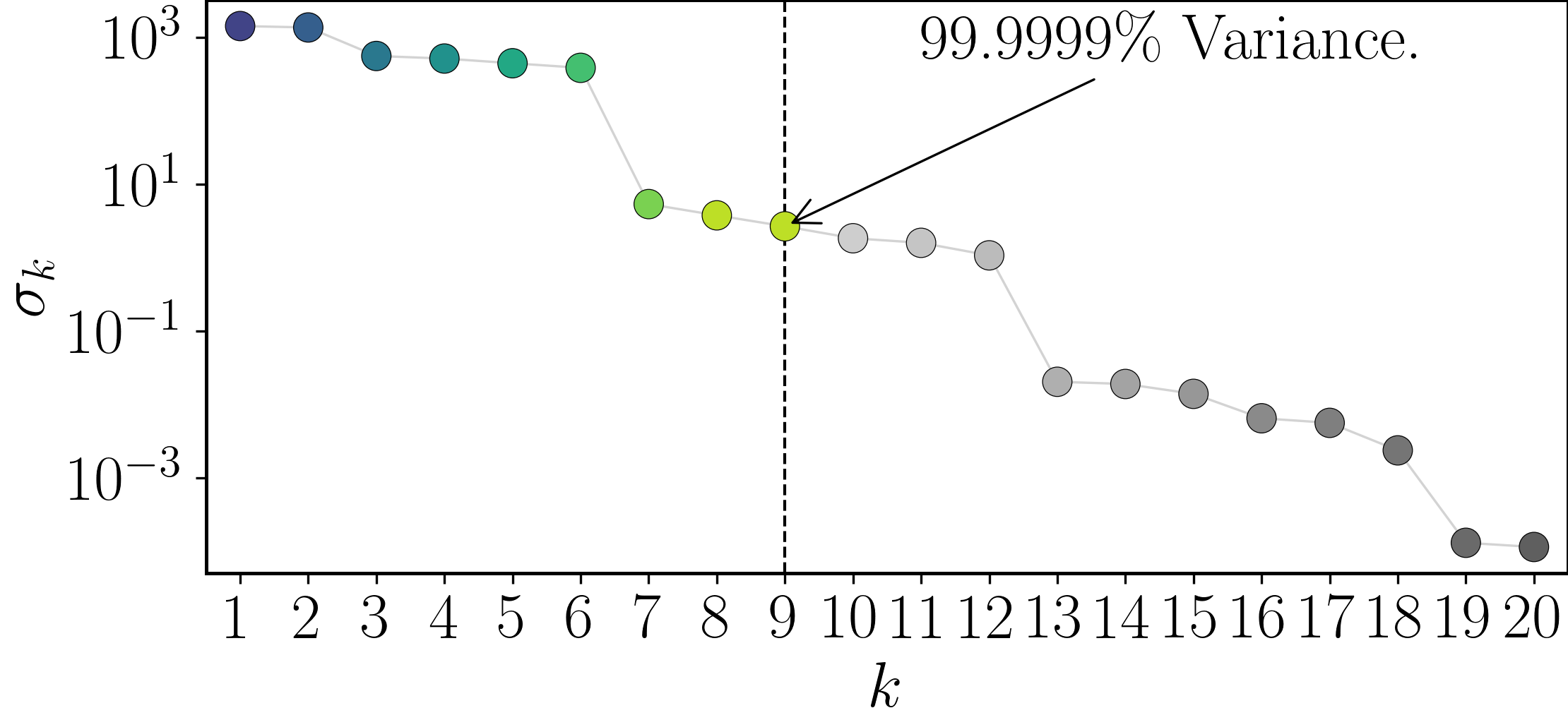}
    \caption{Singular values for the SVD of the block Hankel matrix in Eq.~\eqref{SVD}.}
    \label{fig:variance}
\end{figure}

At first sight, it would seem plausible to use a variance-informed rank selection using Eq.~\eqref{eq:variance_rank_selection_procedure}, since predominant dynamical modes should amount to high percentages of explained variance. However, Fig.~\ref{fig:variance} exhibits how $r=9$, the smallest rank for which \emph{m}HAVOK successfully identified one linear mode, already amounts to more than $99\%$ of the total variance, with adjacent rank values showing similar results. However, this particular value, as well as its neighbors, yields deficient reconstructions, as will be shown in the following (Fig.~\ref{fig:chamfer_r9}). 

Therefore, traditional cutoff criteria based on explained variance are not adequate to work as a proper $r$-selection criterion, motivating a more objective algorithm capable of determining the optimal rank value. 

\subsubsection{\texorpdfstring{$\mathrm{R}^2$-informed quality score}{R²-informed quality score}}
\label{sub_sec:results_quality_score}

\begin{figure}[H]
    \centering
    \includegraphics[width=\linewidth]{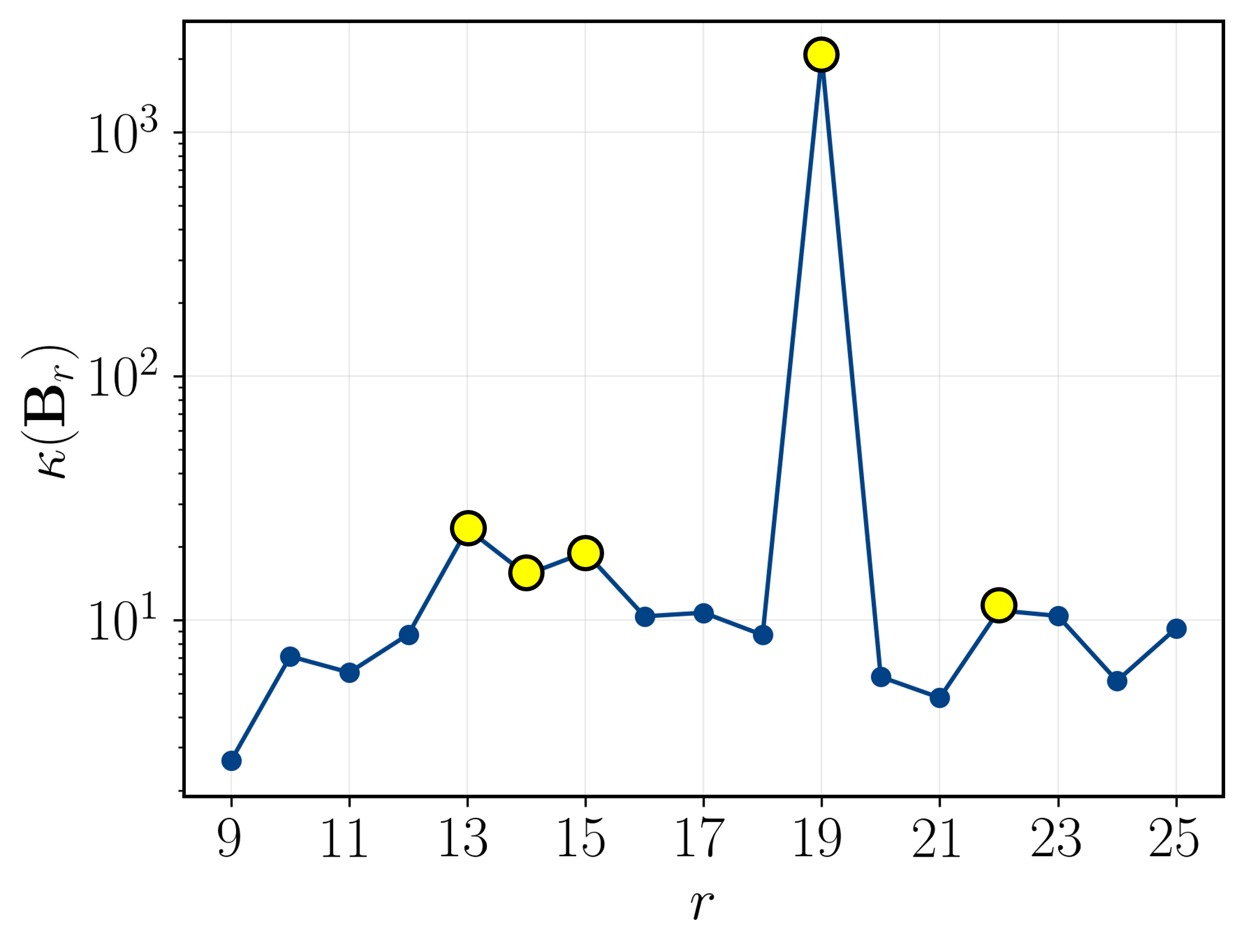}
    \caption{Condition number of $\mathbf{B}_r$ for multiple rank values. The marked values denote ranks with the $30\%$ highest condition number.}
    \label{fig:condition_number}
\end{figure}

For a range of values $r\in[9,25]$, the condition number $\kappa(\mathbf{B}_{r})$ defined in Subsection~\ref{sub_sec:quality_score} was calculated, providing a first filter for the possible $r$-candidates. In general, remarkable peaks in the condition number are connected to optimal reconstructions in multichannel settings. For instance, Fig.~\ref{fig:condition_number} displays a prominent peak at $r=19$, which corresponds to the rank yielding the best coefficient of determination $\overline{R_\text{rec}^2}$. While such peaks provide useful guidance, in practice several candidates need to be examined in order to identify the optimal rank. The following figure highlights the $30\%$ best rank candidates for Sprott system.

% We posit this cutoff rank yields non-redundant, genuine forcing components. To validate this hypothesis, the quality score defined in Eq.~\eqref{eq:quality_score_bn} was calculated for the highest $q=30\%$ condition numbers, confirming that $r_\text{opt}=19$ yields the best coefficient of determination between the original and reconstructed observables. 

\subsection{Quantitative evaluation}\label{sec:sprott_quantitative_evaluation}

\begin{figure}[H]
    \centering
    \begin{subfigure}{\linewidth}
        \centering
        \includegraphics[width=\linewidth]{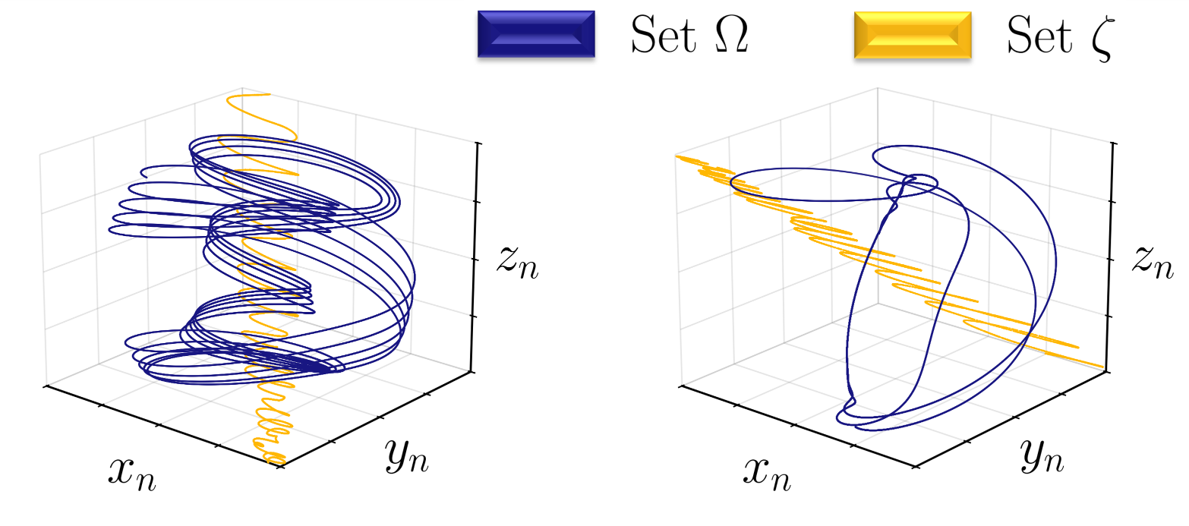}
        \caption{$r=9,\  d_C = 497\times 10^{-3}.$}
        \label{fig:chamfer_r9}
    \end{subfigure}

    \begin{subfigure}{\linewidth}
        \centering
        \includegraphics[width=\linewidth]{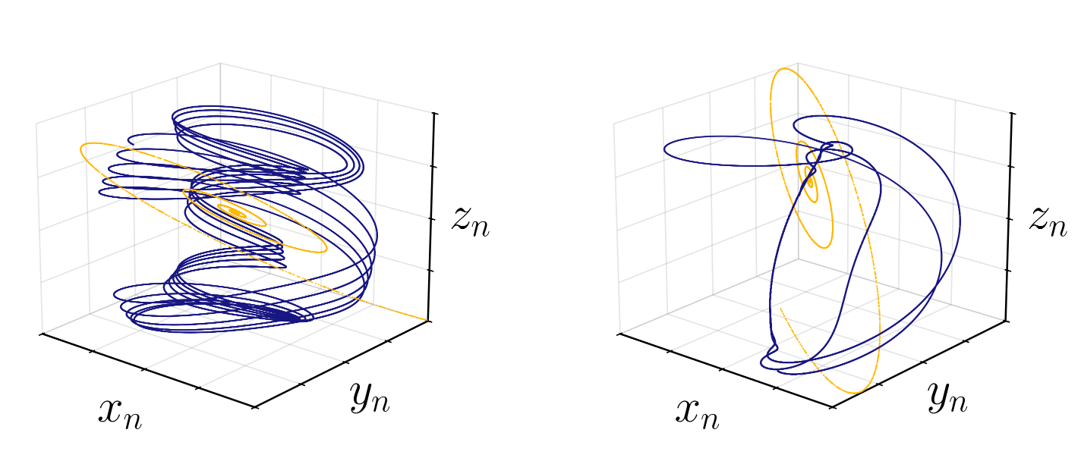}
        \caption{$r=13,\  d_C = 408\times 10^{-3}.$}
        \label{fig:chamfer_r13}
    \end{subfigure}

    \begin{subfigure}{\linewidth}
        \centering
        \includegraphics[width=\linewidth]{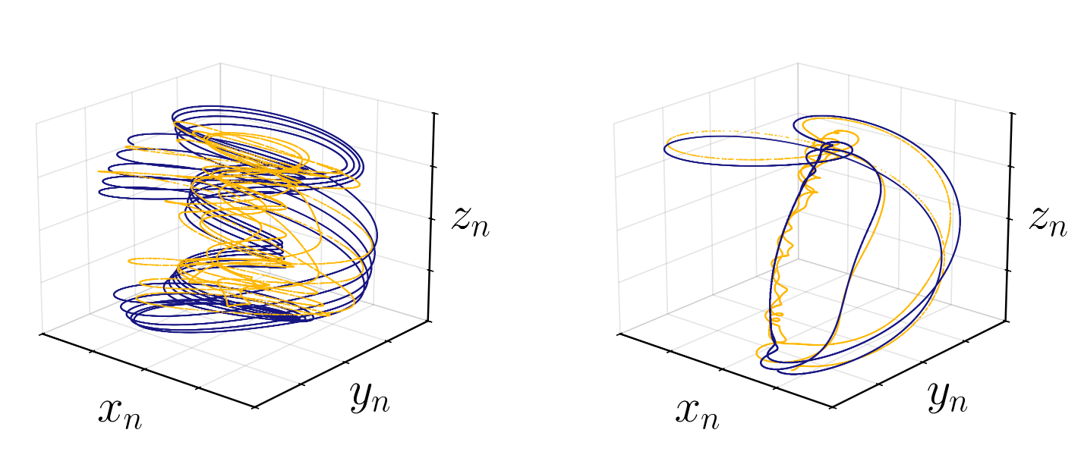}
        \caption{$r=16,\  d_C = 151\times 10^{-3}.$}
        \label{fig:chamfer_r16}
    \end{subfigure}

    \begin{subfigure}{\linewidth}
        \centering
        \includegraphics[width=\linewidth]{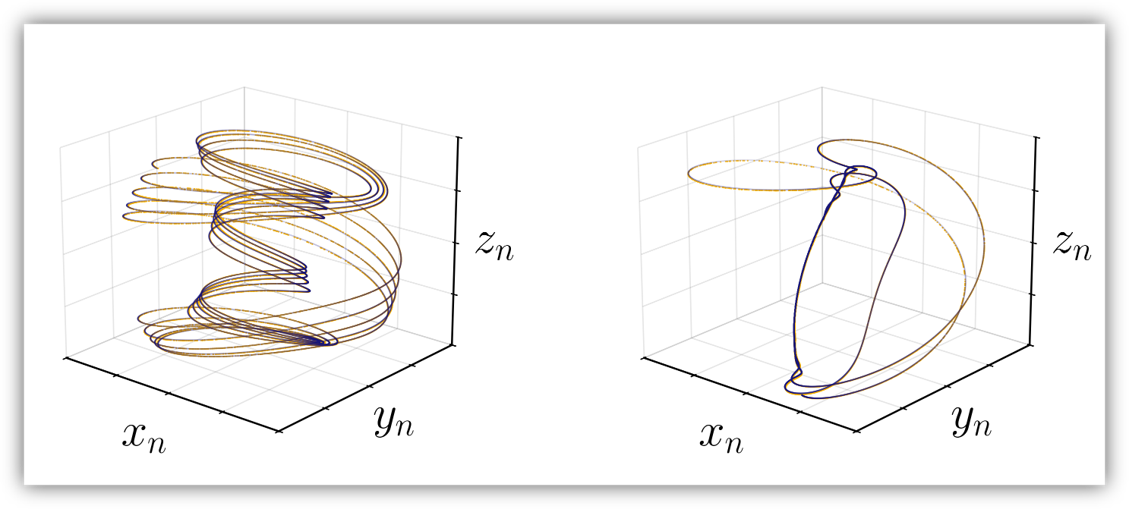}
        \caption{$r_\text{opt}=19,\  d_C = 5\times 10^{-3}.$}
        \label{fig:chamfer_r19}
    \end{subfigure}

    \caption{Comparison of Chamfer distances for different rank values. 
    The sets $\Omega$ and $\zeta$ are defined in Section~\ref{sub_sec:geometric_criteria}.}
    \label{fig:chamfer_cd}
\end{figure}

The Chamfer distance $d_C$ was calculated for some rank values, providing an objective measure of the goodness of the reconstructions. Fig.~\ref{fig:chamfer_cd} demonstrates how $d_C$ tends to decrease as $r$ increases, reaching a minimum at $r_\text{opt}=19$. As opposed to the rank selection in the Lorenz system which is not critical, as already pointed out by Brunton et al. \cite{Brunton_Havok_Nature}, the Sprott system is highly susceptible to the cutoff rank. For instance, $r=20$, one step ahead of the optimal rank value $r_\text{opt}$, yields a distance $d_C=204\times 10^{-3}$, which is almost 40 times greater than $d_C(r_\text{opt})$. Therefore, a slight rank variation generates significantly different reconstructions. 
%Such result should be taken into account when extending \emph{m}HAVOK to a broader range of dynamical systems.

% \begin{figure}[H]
%     \centering
%     \begin{subfigure}{\linewidth}
%         \centering
%         \includegraphics[width=\linewidth]{Images/2_Sprott/chamfer_r9.png}
%         \caption{$r=9,\  d_C = 497\times 10^{-3}.$}
%         \label{fig:chamfer_r9}
%     \end{subfigure}%\hfill
%     \begin{subfigure}{\linewidth}
%         \centering
%         \includegraphics[width=\linewidth]{Images/2_Sprott/chamfer_r13.png}
%         \caption{$r=13,\  d_C = 408\times 10^{-3}.$}
%         \label{fig:chamfer_r13}
%     \end{subfigure}
%         \begin{subfigure}{\linewidth}
%         \centering
%         \includegraphics[width=\linewidth]{Images/2_Sprott/chamfer_r16.png}
%         \caption{$r=16,\  d_C = 151\times 10^{-3}.$}
%         \label{fig:chamfer_r16}
%     \end{subfigure}%\hfill
%         \begin{subfigure}{\linewidth}
%         \centering
%         \includegraphics[width=\linewidth]{Images/2_Sprott/chamfer_r19.png}
%         \caption{$r_\text{opt}=19,\  d_C = 5\times 10^{-3}.$}
%         \label{fig:chamfer_r19}
%     \end{subfigure}
%     \caption{Comparison of Chamfer distances for different rank values. The sets $\Omega$ and $\zeta$ are defined in Section~\ref{sub_sec:geometric_criteria}. }
%     \label{fig:chamfer_cd}
% \end{figure}

% \begin{figure}[H]\ContinuedFloat
%     \centering

%     \begin{subfigure}{\linewidth}
%         \centering
%         \includegraphics[width=\linewidth]{Images/2_Sprott/chamfer_r19.png}
%         \caption{$r_\text{opt}=19,\  d_C = 5\times 10^{-3}.$}
%         \label{fig:chamfer_r19}
%     \end{subfigure}
%     \caption{Comparison of Chamfer distances for different rank values. The sets $\Omega$ and $\zeta$ are defined in Section~\ref{sub_sec:geometric_criteria}. }
%     \label{fig:chamfer_cd}
% \end{figure}

\begin{figure}[H]
    \centering
    \includegraphics[width=\linewidth]{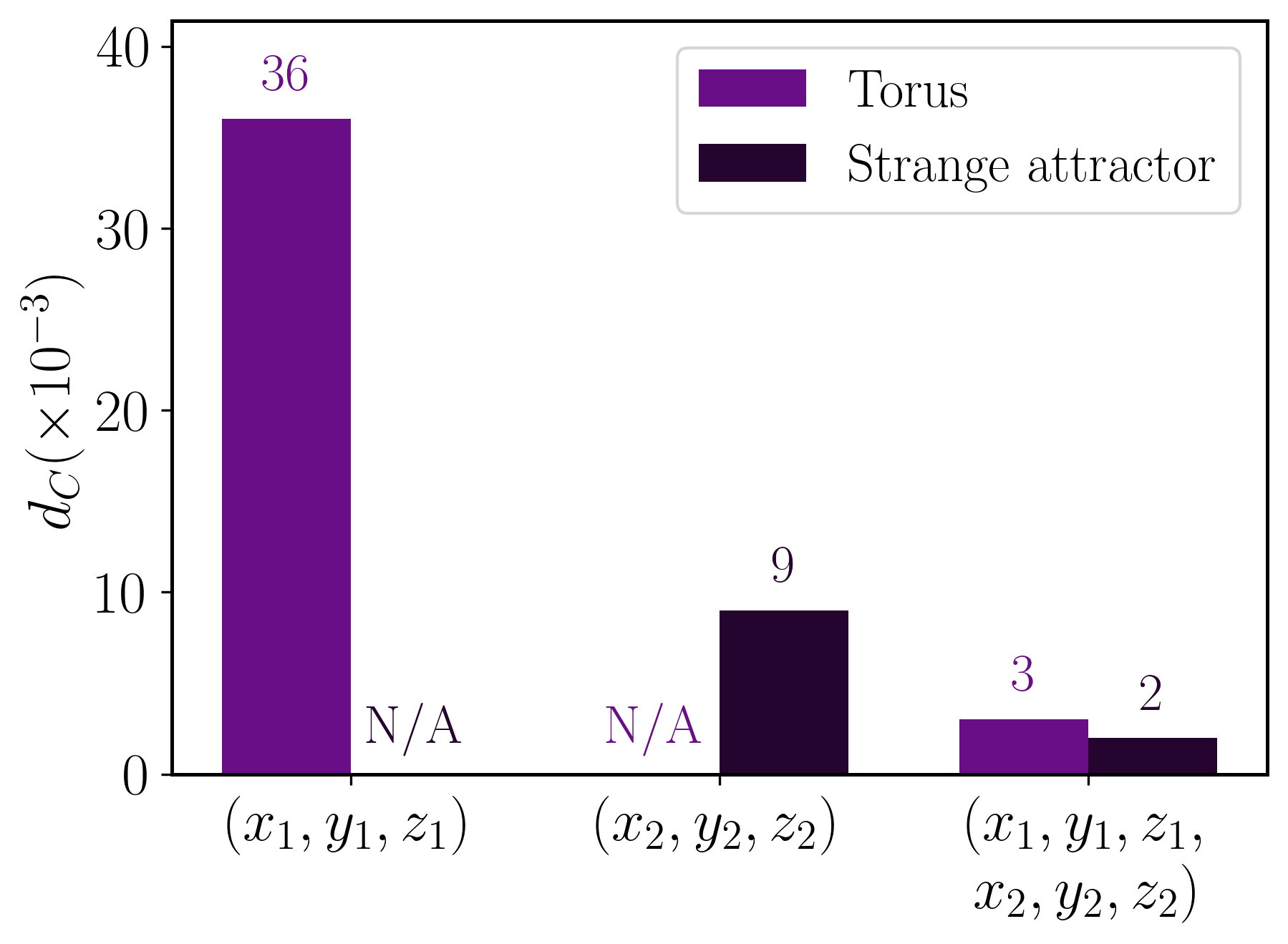}
    \caption{Chamfer distance $d_C$ with respect to the input channels. The simulations were performed using the optimal rank value $r_\text{opt}$ for each case. $(x_1,y_1,z_1)$ and $(x_2,y_2,z_2)$ represent, respectively, the initial conditions from which the torus and strange attractor can be reconstructed.}
    \label{fig:sprott_histogram}
\end{figure}

Interestingly, Fig.~\ref{fig:sprott_histogram} shows how simultaneously providing six channels, $f(\textbf{x}_n)=(x_n,y_n,z_n)$ for both initial conditions (for the invariant torus and the chord-like attractor, respectively), yields a considerable decrease in $d_C$ in comparison to reconstructing the torus and the strange attractor as separate inputs. This can be attributed to the fact that noise correlation is decreased by introducing more observables, as discussed by Deyle and Sugihara \cite{takens_theorem_generalized}.

% r_opt toro = 8
% r_opt cuerda = 9
% r_opt ambos = 19

% In order to measure the goodness of reconstructions, the Chamfer distance defined in Section~\ref{sub_sec:geometric_criteria} was employed. The three channels from both the torus and the strange attractors were provided to \emph{m}HAVOK in separate simulations, and simultaneously, i.e., 6 channels in total, $(x,y,z)$ for each initial condition. 

\section{Conclusions} \label{sec:conclusions} 
\raggedbottom

A generalized multichannel extension of the HAVOK framework (\emph{m}HAVOK) has been developed and validated through simulation. By applying the novel methodology to the Lorenz and Sprott systems, the model's ability to recover the full underlying dynamics is enhanced by incorporating multiple and functionally diverse observables. The algorithm was driven by the key insight that providing the model with more than one informative input signal, alongside a generalized component classification scheme, enriches the embedding and ultimately enables a complete unfolding of the system dynamics. This was particularly evident in the case of the Sprott system, where faithful reconstruction required information from separate initial conditions.

The extension to multichannel embeddings was not only natural but necessary given the practical limitations of real-world measurements, which often consist of indirect or function-based observations. By allowing the integration of such diverse input sources, \emph{m}HAVOK positions itself as a practical and flexible tool. The proposed model is readily applicable in scenarios involving sensor fusion or multiscale dynamics, where capturing the global structure of the attractor is essential for understanding or control.

This framework opens a clear direction for future work by incorporating predictive models for the forcing components. Since \emph{m}HAVOK relies on known or precomputed forcing signals, extending the methodology to forecast those inputs would allow the transition from reconstruction to full forecasting. This would significantly enhance the autonomy and applicability of the model across scientific and engineering domains.

Future work may explore adaptive or data-driven selection of $M$ based on embedding dimension tests, as well as automated threshold techniques. Additionally, a rigorous and well-structured proof of the connection between Koopman's theory and \emph{m}HAVOK is yet to be done and will be discussed in future articles.

\raggedbottom
\newpage

\bibliography{references}

\end{document}